\documentclass[pra,groupedaddress,superscriptaddress,nofootinbib,floatfix,preprintnumbers,longbibliography,notitlepage,tightenlines,11pt]{revtex4-2}
\usepackage[paperwidth=210mm,paperheight=297mm,centering,hmargin=2.3cm,vmargin=2.3cm]{geometry}

\usepackage{amssymb,amsmath,graphicx,color,dsfont}
\usepackage{bm}
\usepackage[tight]{subfigure}
\usepackage[export]{adjustbox}
\usepackage{braket}
\usepackage[colorlinks=true,citecolor=blue,linkcolor=red]{hyperref}
\usepackage[export]{adjustbox}
\usepackage{array}
\newcolumntype{C}[1]{>{\centering\let\newline\\\arraybackslash\hspace{2pt}}m{#1}}
\newcolumntype{R}[1]{>{\centering\let\newline\\\arraybackslash\hspace{0pt}}m{#1}}
\usepackage{multirow}

\begin{document}

\dimen\footins=5\baselineskip\relax

\preprint{\vbox{
\hbox{UMD-PP-021-07,
INT-PUB-21-030}
}}

\title{On the Extraction of Low-energy Constants of Single- and Double-$\beta$ Decays 
\\
from Lattice QCD: A Sensitivity Analysis
}
\author{Zohreh Davoudi{\footnote{\tt davoudi@umd.edu}}
}
\affiliation{Maryland Center for Fundamental Physics and Department of Physics, 
University of Maryland, College Park, MD 20742, USA}

\author{Saurabh V. Kadam{\footnote{\tt ksaurabh@umd.edu }}
}
\affiliation{Maryland Center for Fundamental Physics and Department of Physics, 
University of Maryland, College Park, MD 20742, USA}

\begin{abstract} 
Lattice quantum chromodynamics (LQCD) has the promise of constraining low-energy constants (LECs) of nuclear effective field theories (EFTs) from first-principles calculations that incorporate the dynamics of quarks and gluons. Given the Euclidean and finite-volume nature of LQCD outputs, complex mappings are developed in recent years to obtain the Minkowski and infinite-volume counterparts of LQCD observables. In particular, as LQCD is moving toward computing a set of important few-nucleon matrix elements at the physical values of the quark masses, it is important to investigate whether the anticipated precision of LQCD spectra and matrix elements will be sufficient to guarantee tighter constraints on the relevant LECs than those already obtained from phenomenology, considering the non-trivial mappings involved. 
With a focus on the leading-order LECs of the pionless EFT, $L_{1,A}$ and $g_{\nu}^{NN}$, which parametrize, respectively, the strength of the isovector axial two-body current in a single-$\beta$ decay (and other related processes such $pp$ fusion), and of the isotensor contact two-body operator in the neutrinoless double-$\beta$ decay within the light neutrino exchange scenario, the expected uncertainty on future extractions of $L_{1,A}$ and $g_{\nu}^{NN}$ are examined using synthetic data at the physical values of the quark masses. It is observed that achieving small uncertainties in $L_{1,A}$ will be challenging, and (sub)percent-level precision in the two-nucleon spectra and matrix elements is essential in reducing the uncertainty on this LEC compared to the existing constraints. On the other hand, the short-distance coupling of the neutrinoless double-$\beta$ decay, $g_{\nu}^{NN}$, is shown to be less sensitive to uncertainties on both LQCD energies and the matrix element, and can likely be constrained with percent-level precision in the upcoming LQCD calculations.

\end{abstract}

{\let\newpage\relax\maketitle}


\section{Introduction
\label{sec: Introduction}
}
\noindent
Nuclear reactions mediated by weak interactions are central to a variety of frontier problems in nuclear and astrophysics as well as high-energy physics. Single-weak-current processes like $pp$-fusion and (anti)neutrino-deuteron scattering are two prominent examples. The former is a critical process in understanding the energy production mechanism in a range of stars~\cite{Adelberger:2010qa}, and the latter is used to probe properties of neutrinos in several neutrino experiments~\cite{Aharmim:2011vm,Fukuda:2001nj,Fukuda:2002pe}. At the next order in weak currents, double-$\beta$ decay transitions are of major importance. Two important modes of this transition are two-neutrino double-$\beta$ ($2\nu\beta\beta$) decay and neutrinoless double-$\beta$ ($0\nu\beta\beta$) decay. The former process conserves the total lepton number~\cite{GoeppertMayer:1935qp}, and is the rarest Standard Model (SM) process that has been measured~\cite{Barabash:2020nck}. Besides providing insights into the SM weak interactions and nuclear structure, $2\nu\beta\beta$ decay can also shed light on potential beyond SM scenarios~\cite{Deppisch:2020mxv}. The $0\nu\beta\beta$ mode is forbidden in the SM as it changes the lepton number by two units, and if observed, would indicate that neutrinos are of Majorana type~\cite{Schechter:1981bd}. An extensive experimental program continues to seek evidence for $0\nu\beta\beta$ decays~\cite{Bilenky:2014uka, DellOro:2016tmg,Biassoni:2020byh,Dolinski:2019nrj,Cappuzzello:2018wek,Cappuzzello:2016zlj, Bilenky:2014uka,Bilenky:2020wjn}. However, the new-physics implications of the current and the future double-$\beta$ decay measurements are limited by the uncertainties in the theoretical predictions of their decay rates.\par

A major source of uncertainty in calculating the decay rate of weak processes is the matrix elements (MEs) of weak currents between the initial and final nuclear states. For energies well below the pion mass, $m_{\pi}$, that is often relevant for single-weak processes in the few-nucleon sector, pionless EFT~\cite{Kaplan:1998tg,Kaplan:1998we,vanKolck:1998bw,Bedaque:1997qi,Bedaque:1998mb,Bedaque:1999vb,Chen:1999tn} accurately describes the dynamics, see Ref.~\cite{Hammer:2019poc} for a review. For the double-$\beta$ decays that naturally occur in large nuclear isotopes, the corresponding nuclear-ME calculations suffer from uncertainties that stem from both approximations in quantum many-body methods as well as uncertainties in (multi)nucleon interactions and weak currents~\cite{Engel:2016xgb,Giuliani:2012zu}. The latter can be mitigated by improving the accuracy of MEs in the two-nucleon (NN) sector using an effective Lagrangian along with a power-counting scheme, and then using them as an input in an \textit{ab initio} framework to calculate the many-body MEs for larger nuclei~\cite{Coraggio:2020iht,Engel:2016xgb}. The NN transitions between the two-neutron initial state, $nn$, and the two-proton final state, $pp$, are not observed in free space, but they occur as off-shell subprocesses in transitions of larger nuclei. The typical Fermi momentum of nucleons in these nuclei is comparable to $m_{\pi}$, but at a first approximation, the pionless EFT is expected to provide a good description. Subsequently, the effect of pions can be included systematically using pionfull EFT~\cite{Kaplan:1998tg,Kaplan:1998we} or chiral nuclear EFTs~\cite{Weinberg:1990rz, Weinberg:1991um, Machleidt:2011zz}.\par

For SM processes involving more than two nucleons, the nuclear MEs of isovector axial-vector currents corresponding to Gamow-Teller transitions are not constrained precisely in pionless EFT. This is in part due to a large uncertainty on the renormalization-scale ($\mu$) dependent LEC $L_{1,A}$ that contributes at the next-to-leading order (NLO) and determines the strength of the momentum-independent isovector axial-vector two-body current~\cite{Kong:1999tw,Butler:1999sv,Butler:2000zp}. While constituting only a few percent of the total amplitude, the contribution to the Gamow-Teller transitions from the $L_{1,A}$ term remains the dominant source of uncertainty in determining the decay rate of processes such as $pp$ fusion in Sun and similar stars~\cite{Adelberger:2010qa}. The value of $L_{1,A}$ determined from experimental data has improved over the years~\cite{Chen:2002pv, Butler:2002cw, Butler:2000zp, Chen:2005ak, De-Leon:2016wyu}, with the most recent constraint given by\footnote{Throughout this work, values of $\mu$-dependent LECs are given at $\mu=m_{\pi}$.} $L_{1,A}= 4.9 ^{+1.9}_{-1.5}\text{ fm}^3$~\cite{Acharya:2019fij}, which has a significant uncertainty.  On the other hand, no experimental constraint exists on the nuclear ME of $0\nu\beta\beta$ decay transition due to lack of observation. Furthermore, recent analyses in the light neutrino exchange scenario of the $0\nu\beta\beta$ decay transition in the two-nucleon sector, i.e. $nn \to pp e^- e^-$, have shown that the corresponding nuclear ME is unknown even at the leading order (LO) in pionless EFT~\cite{Cirigliano:2017tvr,Cirigliano:2018hja,Cirigliano:2019vdj}. In fact, a new $\mu$-dependent LEC, $g_\nu^{NN}$, is needed at LO for the decay amplitude to be manifestly renormalizable. Recently, an indirect estimate of $g_\nu^{NN}$ was obtained in Refs.~\cite{Cirigliano:2020dmx,Cirigliano:2021qko}: $\widetilde{g}_{\nu}^{NN} = 1.3 \pm 0.6$,
where $\widetilde{g}_{\nu}^{NN}$ is a dimensionless parameter related to $g_{\nu}^{NN}$. Subsequent analyses using this value showed that the missing short-range contribution to the nuclear ME of various candidate nuclei is comparable to the rest of the contributions~\cite{Wirth:2021pij,Jokiniemi:2021qqv}. This indicates the importance of improving the constraint on $g_\nu^{NN}$, preferably using a  first-principles approach such as LQCD.\par

%
A direct way of constraining nuclear MEs is to solve the underlying short-distance theory of quantum chromodynamics (QCD) numerically using the technique of LQCD~\cite{Davoudi:2020ngi,Briceno:2014tqa,Cirigliano:2019jig,Kronfeld:2019nfb,Drischler:2019xuo,Cirigliano:2020yhp}. LQCD was in fact used  in Ref.~\cite{Savage:2016kon} to constrain $L_{1,A}$ from the relevant LQCD three-point correlation functions albeit at unphysical quark masses corresponding to $m_{\pi}\approx 806$ MeV, see also Ref.~\cite{Detmold:2021oro}. The obtained value of $L_{1,A}=3.9(1.4) \text{ fm}^3$ at the physical quark masses required an uncertain quark-mass extrapolations but found to be comparable to experimental constraints with similar uncertainties. On the other hand, no LQCD determination of the $g_\nu^{NN}$ coupling is yet reported although progress in simpler $0\nu\beta\beta$ processes in the pion sector is being made in recent years~\cite{Feng:2018pdq, Tuo:2019bue, Detmold:2020jqv, Nicholson:2018mwc}. In LQCD, the QCD action is defined on a finite spacetime grid with a Euclidean time, and the $n$-point correlation functions are computed using Monte Carlo methods. A formalism for obtaining two-hadron scattering amplitudes from finite-volume (FV) Euclidean correlation functions was introduced by L\"uscher~\cite{Luscher:1986pf,Luscher:1990ux} and extended to other systems in Refs.~\cite{Rummukainen:1995vs, Beane:2003da, Kim:2005gf, He:2005ey, Davoudi:2011md, Leskovec:2012gb, Briceno:2012yi, Hansen:2012tf, Gockeler:2012yj, Briceno:2013lba, Feng:2004ua, Lee:2017igf, Bedaque:2004kc, Luu:2011ep, Briceno:2013hya, Briceno:2013bda, Briceno:2014oea,Polejaeva:2012ut,Briceno:2012rv, Hansen:2014eka, Hansen:2015zga, Hammer:2017uqm,Hammer:2017kms,Guo:2017ism,Mai:2017bge}. The formalism for obtaining transition amplitudes of processes involving external currents was first developed by Lellouch and L\"uscher~\cite{Lellouch:2000pv} and later generalized in Refs.~\cite{Briceno:2012yi,Christ:2005gi,Meyer:2011um,Bernard:2012bi,Beane:2014qha,Detmold:2004qn, Meyer:2011um, Briceno:2012yi, Bernard:2012bi, Briceno:2014uqa, Feng:2014gba, Briceno:2015csa, Briceno:2015tza, Hansen:2021ofl}. For the hadronic MEs involving long-range processes, the generalization of the above mappings resolves an additional complexity arising from the relative time between the two hadronic currents~\cite{Shanahan:2017bgi,Tiburzi:2017iux,Feng:2018pdq,Tuo:2019bue,Briceno:2019opb,Detmold:2020jqv,Feng:2020nqj,Davoudi:2020xdv,Davoudi:2020gxs,Christ:2012se,Christ:2020hwe}. Recently, we applied this formalism to single- and double-$\beta$ decays in the NN sector to obtain the needed matching relations that constrain the $L_{1,A}$ and $g_\nu^{NN}$ LECs from the LQCD output~\cite{Davoudi:2020xdv,Davoudi:2020gxs}.\par

Given the complexity of the matching relations involved, it is not immediately obvious what the precision requirement of the upcoming LQCD studies at the physical quark masses should be to reach the precision goal of the LECs, that is to be compatible or superior to phenomenological constraints. In particular, it is important to ask if anticipated uncertainties on the lowest-lying FV energies and on the MEs, as well as achievable physical volumes in LQCD, will guarantee precise determinations of LECs such as $L_{1,A}$ and $g_\nu^{NN}$. As a result, in this paper we embark on an investigation based on synthetic data to determine the sensitivity of the output of the matching relations (hence the LECs) along with their uncertainties on the values and uncertainties of the input to these relations, namely the LQCD energies and MEs. This also allows determining the range of volumes which leads to better constraints, hence guiding future LQCD calculations on their resource planning. This follows the spirit of Ref.~\cite{Briceno:2013bda} which demonstrated that a precise determination of the small S-D mixing parameter in the deuteron channel from LQCD is achievable in future LQCD calculations of the lowest-lying spectra of NN systems in boosted frames. This investigation, furthermore, aligns with recent valuable analyses of the sensitivity of nuclear spectra and MEs to the uncertainties in the input LECs of interactions and currents, when those uncertainties are propagated through \emph{ab initio} many-body calculations~\cite{Ekstrom:2019lss}.
\begin{figure}[t]
\centering
\includegraphics[scale=0.9]{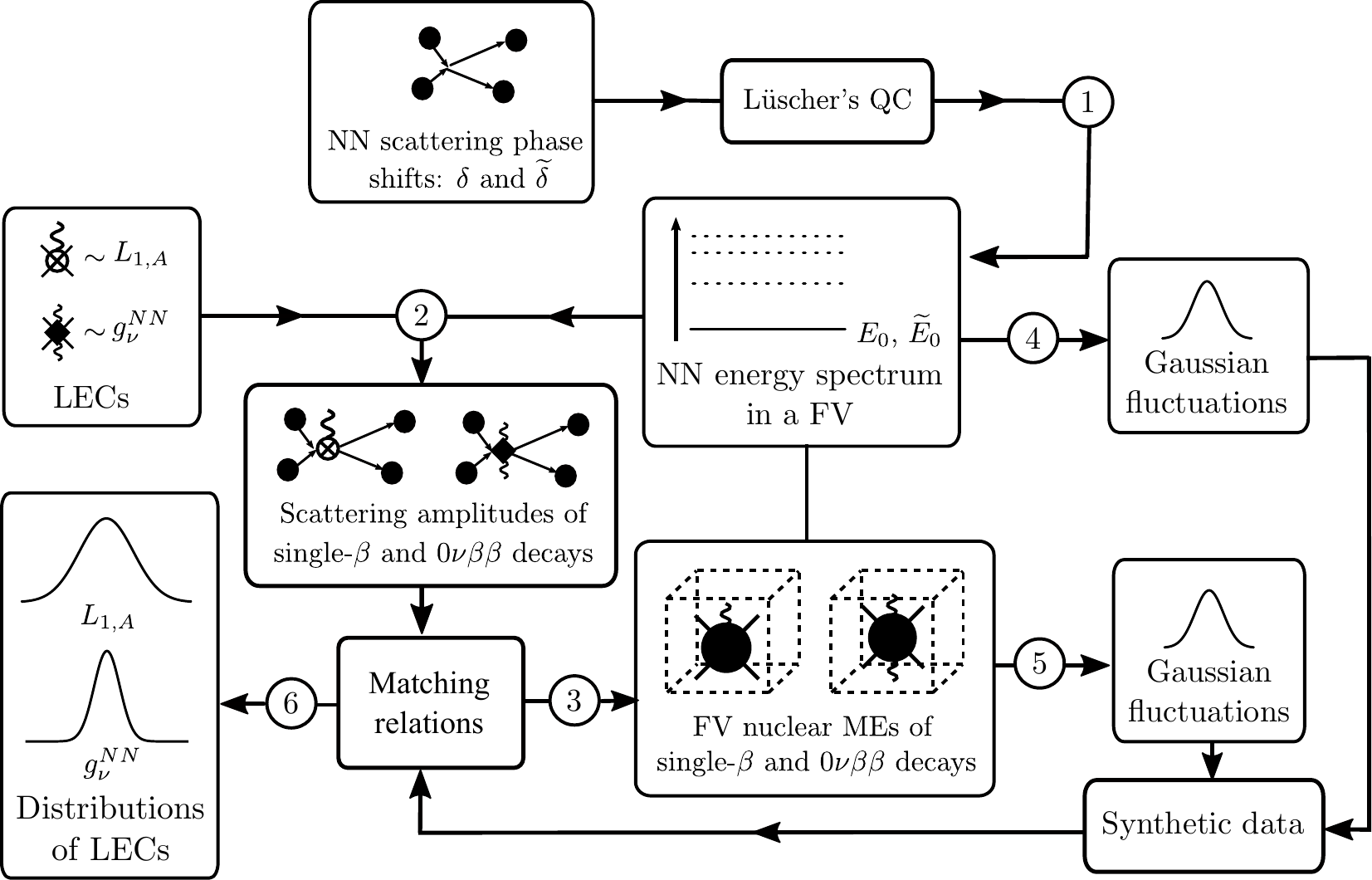}
\caption{The procedure used to perform the sensitivity analysis of $L_{1,A}$ and $g_\nu^{NN}$. The sequence of steps followed is indicated by the numbers enclosed in the circles. The LECs $L_{1,A}$ and $g_\nu^{NN}$ are represented by a crossed circle and a solid diamond, respectively. The wavy line denotes external leptons from a single weak-current insertion. A nucleon is denoted by the small solid circle in the diagrams for NN processes in infinite volume. Dotted lines in the NN energy spectrum in a finite volume are the excited-state energies, and the ground state energy, $E_0$ ($\widetilde{E}_0$), in the spin-singlet (spin-triplet) channel is denoted by the solid line. The FV nuclear MEs for the decay transitions are represented by large solid circles enclosed in dotted cubes with one and two weak-current insertions, respectively. The solid line denotes the FV nucleon state. The simulation of LQCD uncertainties using Gaussian fluctuations and uncertainty analysis of LECs from the synthetic data is discussed in Secs.~\ref{sec: L1A} and~\ref{sec: gvNN}. \label{fig: flowchart}}
\end{figure}

Explicitly, we consider the single-$\beta$ decay (Sec.~\ref{sec: L1A}) and $0\nu\beta\beta$ decay (Sec.~\ref{sec: gvNN}) transitions in the two-nucleon sector: $nn \to np e^-\bar{\nu}_e$ and $nn \to pp e^-e^-$, respectively. First using the L\"uscher's quantization condition (QC), the low-energy spectra of NN systems in a range of spatial cubic volumes with periodic boundary conditions (PBCs) are calculated using the phase shifts reported in the experimental NN scattering database~\cite{NNonline}. These spectra are expected to be the same as those calculated from the two-point function with LQCD at the physical quark masses, up to exponential corrections in $m_\pi$. Second, the central values of $L_{1,A}$ and $g_{\nu}^{NN}$ from Ref.~\cite{Acharya:2019fij} and Refs.~\cite{Cirigliano:2020dmx,Cirigliano:2021qko} are used to evaluate the physical transition amplitudes for single- and (neutrinoless) double-$\beta$ decay processes with initial and final energies set to the lowest energy eigenvalues obtained in the first step. These scattering amplitudes are then used in matching relations of Refs.~\cite{Davoudi:2020xdv} and \cite{Davoudi:2020gxs}, respectively, to obtain a reasonable guess for the central values of the corresponding FV three- and four-point functions. Next, Gaussian fluctuations are introduced to the quantities that are expected to be extracted from LQCD, namely the FV energy eigenvalues and the three- and four-point functions, to generate a set of synthetic data for performing the sensitivity analysis. This introduces uncertainties in the supposedly LQCD ingredients. Finally, matching relations are used once again to obtain $L_{1,A}$ and $g_{\nu}^{NN}$ from the synthetic dataset, along with their uncertainties. Figure~\ref{fig: flowchart} summarizes the procedure used for performing the sensitivity analysis of this work. \par

A detailed account of our findings is provided in Sec.~\ref{sec: conclusions}. To summarize, achieving small uncertainties in $L_{1,A}$ is found to be more challenging than $g_{\nu}^{NN}$, and demands (sub)percent-level precision in the two-nucleon spectra and the ME to supersede the current phenomenological constraints. On the other hand, the short-distance coupling of the neutrinoless double-$\beta$ decay, $g_{\nu}^{NN}$, turns out to be less sensitive to uncertainties on both LQCD energies and the ME, and promises competitive precision compared with the current indirect estimates, even with few-percent uncertainties on LQCD energies and MEs. The volume requirements are moderate and for ground-state to ground-state transitions, smaller volumes are shown to lead to more precise extractions.

\section{Formalism
\label{sec: Formalism}
}
\noindent
In this section, we present a brief overview of the pionless EFT~\cite{Kaplan:1998tg,Kaplan:1998we,vanKolck:1998bw,Bedaque:1997qi,Bedaque:1998mb,Bedaque:1999vb,Chen:1999tn} employed to evaluate the hadronic scattering amplitudes of $nn \to np e^-\bar{\nu}_e$ and $nn \to pp e^-e^-$ transitions. Furthermore, the FV ingredients required to perform the sensitivity analyses of Secs.~\ref{sec: L1A} and~\ref{sec: gvNN} are obtained via the application of L\"uscher's quantization condition that relates the FV energy eigenvalues to the physical two-hadron scattering amplitudes. Our notation follows that used in Ref.~\cite{Davoudi:2020xdv}.

\subsection{Pionless EFT
\label{subec: Pionless EFT}}
In the pionless EFT, the hadronic Lagrangian is arranged according to the number of nucleons. The relativistic corrections and the isospin-breaking effects contribute at higher orders than considered in this work. The single-nucleon Lagrangian is given by
\begin{equation}
	{\cal L}_{(1)}=N^{\dagger }\bigg(i\partial_{t}+\frac{{\nabla}^2}{2M}\bigg)N+\cdots,
	\label{eq: EFT 1 nucleon Lagrangian}
\end{equation}
where ellipsis denotes relativistic corrections. Here, $\partial_t$ is the time derivative and ${\bf \nabla}$ is the spatial gradient operator. $N=(p,n)^T$ is an isospin doublet composed of the proton, $p$, and the neutron, $n$, fields, each with mass $M$. The NN contact interactions are governed by the Lagrangian,
\begin{align}
	{\cal L}_{(2)}=&-C_{0}(N^{T}\mathcal{P}_iN)^{\dagger }(N^{T}\mathcal{P}_iN)  \nonumber -\widetilde{C}_{0}(N^{T}\widetilde{\mathcal{P}}_iN)^{\dagger }(N^{T}\widetilde{\mathcal{P}}_iN)\,+ \nonumber \\
	&~~~{\frac{C_{2}}{8}}\left[ (N^{T}\mathcal{P}_i%
	N)^{\dagger }(N^{T}(\overleftarrow{{\bf\nabla}}^{2}\mathcal{P}_i-2%
	\overleftarrow{{\bf\nabla}}\cdot \mathcal{P}_i\overrightarrow{{\bf\nabla}}+%
	\mathcal{P}_i\overrightarrow{{\bf\nabla}}^{2})N)+{\rm H.c.}\right]+
	\nonumber\\
	&~~~{\frac{\widetilde{C}_{2}}{8}}\left[ (N^{T}\widetilde{\mathcal{P}}_iN)^{\dagger }(N^{T}(%
	\overleftarrow{{\bf\nabla}}^{2}\widetilde{\mathcal{P}}_i-2\overleftarrow{{\bf\nabla}}\cdot \widetilde{\mathcal{P}}_i%
	\overrightarrow{{\bf\nabla}}+\widetilde{\mathcal{P}}_i\overrightarrow{{\bf\nabla}}^{2})N)+{\rm H.c.}\right]
	 + \cdots ,
	 \label{eq: Nucleon EFT Lagrangian}
\end{align}
The overhead arrow indicates which nucleon field is acted by the derivative operator, and ellipsis denotes higher-derivative operators that will not contribute to the order at which the analysis of this work is performed. Index $i=1,2,3$ is summed over. $\mathcal{P}_i$ and $\widetilde{\mathcal{P}}_i$ are the spin-isospin projection operators for the spin-singlet $(^1S_0)$ and spin-triplet $(^3S_1)$ channels, respectively.\footnote{The spin-triplet channel couples S and D partial waves. Since partial-wave mixing both in infinite and finite volumes is neglected in this work, the spin-triplet channel will be denoted by $^3S_1$ instead of $^3S_1-{^3}D_1$.} Strong-interaction LECs for these channels are distinguished by an overhead tilde for the $^3S_1$ channel.\par

For NN systems in the $^1S_0$ channel at a low center-of-mass (CM) energy, $E$, the scattering amplitude, $\mathcal{M}$ is described by an $S$-wave scattering phase shift, $\delta$,
\begin{equation}
    \mathcal{M} = \frac{4\pi}{M} \frac{1}{p\cot{\delta}-i p} ,
    \label{eq: scattering amplitude}
\end{equation}
where $p=\sqrt{ME}$ and higher partial-wave contributions are ignored. Below the t-channel cut, the effective-range function $p\cot{\delta}$ can be expansed in $p^2$ near $p^2=0$, resulting in an effective-range expansion, 
\begin{equation}
    p\cot{\delta} = -\frac{1}{a} + \frac{1}{2} r p^2  + \cdots,
    \label{eq: ERE}
\end{equation}
where $a$ is the scattering length, $r$ is the effective range, and ellipsis denotes higher-order terms that will be neglected in this analysis. In the pionless EFT with the Kaplan-Savage-Wise power counting~\cite{Kaplan:1998tg,Kaplan:1998we}, the S-wave scattering amplitude is expanded to LO and NLO  amplitudes:
\begin{align}
     \mathcal{M}^{\rm (LO)} &=
    -\frac{4\pi}{M} \frac{1}{(1/a + ip)},
    \label{eq: MLO NN scattering}
    \\
    \mathcal{M}^{\rm (NLO)} &=
    -\frac{2\pi}{M} \frac{r\,p^2}{(1/a + ip)^2}.
    \label{eq: MNLO NN scattering}
\end{align}
The LO amplitude, $\mathcal{M}^{\rm (LO)}$, is given by the tree-level NN contact interaction, $C_0$, plus any number of $C_0$ vertices connected by the s-channel two-nucleon loops. The NLO amplitude,  $\mathcal{M}^{\rm (NLO)}$, involves one insertion of the NN derivative coupling, $C_2$, dressed by the NN propagator and the LO amplitude from both sides. The NN s-channel loop is an ultraviolet (UV) divergent integral that is regularized with the power-divergence subtraction scheme introduced in Ref.~\cite{Kaplan:1998tg}. By comparing these amplitudes with Eqs.~\eqref{eq: MLO NN scattering} and~\eqref{eq: MNLO NN scattering} for the $^1S_0$ channel, the NN contact interactions at a given renormalization scale, $\mu$, can be expressed in terms of the effective-range expansion parameters defined in Eq.~\eqref{eq: ERE}:
\begin{align}
    C_0(\mu) & = \frac{4\pi}{M} \frac{1}{(-\mu + 1/a)},
    \label{eq: C0 scale relation}\\
    C_2(\mu) & = \frac{2\pi}{M} \frac{r}{(-\mu + 1/a)^2}.
    \label{eq: C2 scale relation}
\end{align}
Equations~\eqref{eq: scattering amplitude}-\eqref{eq: C2 scale relation} are valid for the $^3S_1$ channel too upon replacements, $\delta \to \widetilde{\delta}$,\footnote{$\tilde{\delta}$ is the $\alpha$-wave phase shift in the Blatt-Biedenharn parametrization of the coupled $^3S_1-{^3}D_1$ channel~\cite{blatt1952neutron}, but here it will be referred to as an S-wave phase shift for simplicity.} $a \to \widetilde{a}$, $r \to \widetilde{r}$, $\mathcal{M}^{\rm (LO)} \to \widetilde{\mathcal{M}}^{\rm (LO)}$, $\mathcal{M}^{\rm (NLO)} \to \widetilde{\mathcal{M}}^{\rm (NLO)}$, $C_0 \to \widetilde{C}_0$ and $C_2 \to \widetilde{C}_2$, where the overhead tilde denotes the analogous quantity in the $^3S_1$ channel.\footnote{Overhead tilde is used throughout to denote two-nucleon quantities in the $^3S_1$ channel. The only exceptions to this rule are $\widetilde{L}_{1,A}$ and $\widetilde{g}_{\nu}^{NN}$ that denote renormalization-scale independent LECs in  Secs.~\ref{sec: L1A} and~\ref{sec: gvNN}. The convention for these LECs is maintained to be consistent with the literature.} The NN scattering amplitudes and contact LECs introduced in this section are needed in matching relations for $L_{1,A}$ and $g_{\nu}^{NN}$ in Eq.~\eqref{eq: matching relation for L1A} and~\eqref{eq: matching relation gvNN}, respectively.\par
%

\subsection{L\"uscher's method
\label{subec: FV formalism}}
In LQCD, the $n$-point correlation functions are computed on a finite Euclidean spacetime lattice. Assuming the continuum limit for a hypercubic lattice with periodic boundary conditions, L\"uscher's quantization condition gives a direct relation between the FV energy eigenvalues of two hadrons obtained from LQCD and the corresponding scattering amplitudes. The mapping is valid up to exponentially suppressed corrections governed by the range of the interactions. For the low-energy NN systems, the interaction range is set by the Compton wavelength of the pion. The quantization conditions are then valid up to $\mathcal{O}(e^{-m_\pi L})$ corrections, where $L$ denotes the spatial extent of the volume. \par
\begin{figure}[t]
    \centering
    \includegraphics[scale=1]{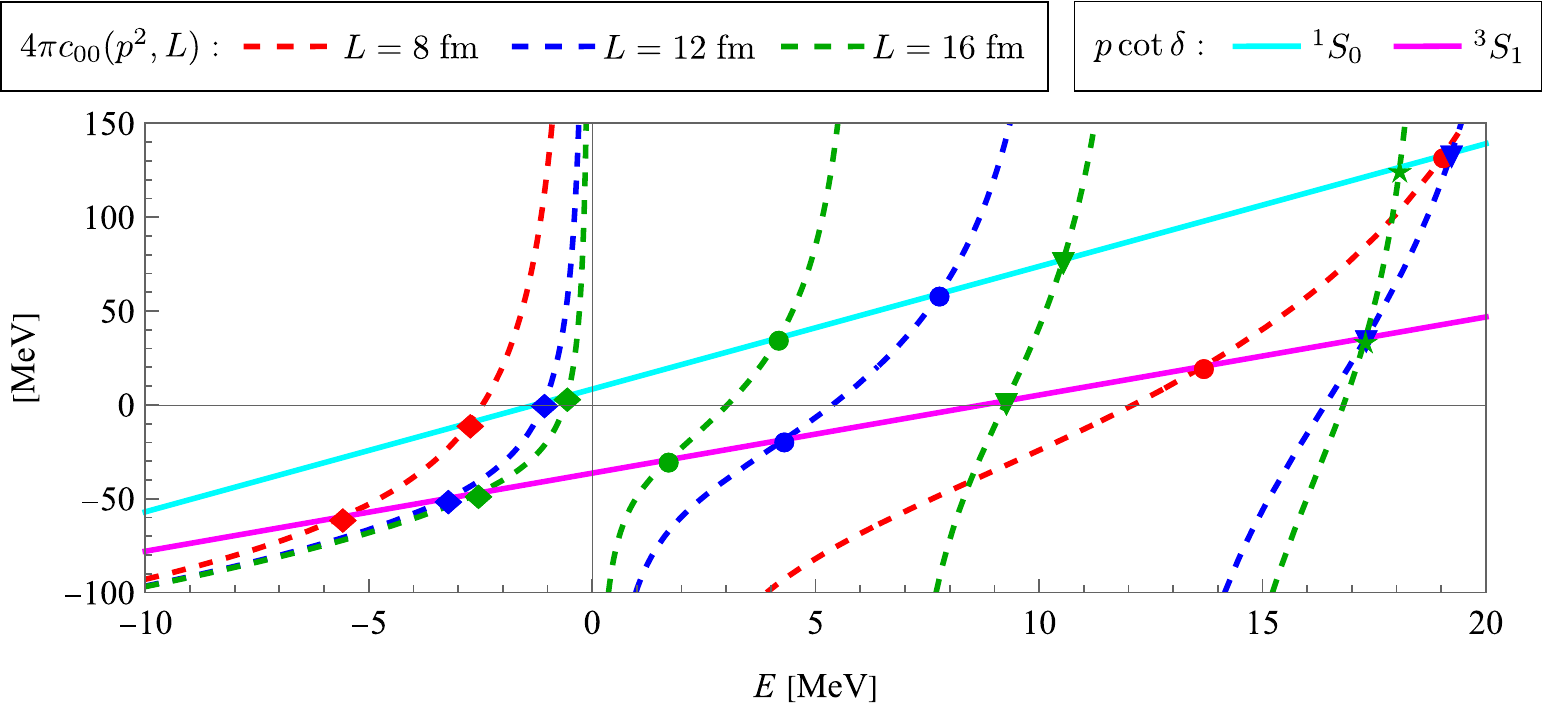}
    \caption{The effective-range function (solid lines) and L\"uscher's function (dotted lines) in Eq.~\eqref{eq: quantization condition} are plotted independently against the CM energy of NN systems. Equation~\eqref{eq: ERE} is used for the effective-range function with the effective-range expansion parameters given in Eq.~\eqref{eq: ERE values} for the two channels, $^1S_0$ (cyan) and $^3S_1$ (magenta). The function $4\pi c_{00}(p^2,L)$ is plotted for three different volumes with $L=8\;{\rm fm}$ (red), $L=12\;{\rm fm}$ (blue) and $L=16\;{\rm fm}$ (green). The diamonds, circles, triangles, and stars denote, respectively, the location of energy eigenvalues of the ground, first, second, and third excited states in each volume, and satisfy the quantization condition in Eq.~\eqref{eq: quantization condition} (and its counterpart for the $^3S_1$ channel). The numerical values associated with this figure are provided in Appendix~\ref{app:detail}.
    \label{fig: QC plot}}
\end{figure}
\begin{figure}[t]
    \centering
    \includegraphics[scale=1]{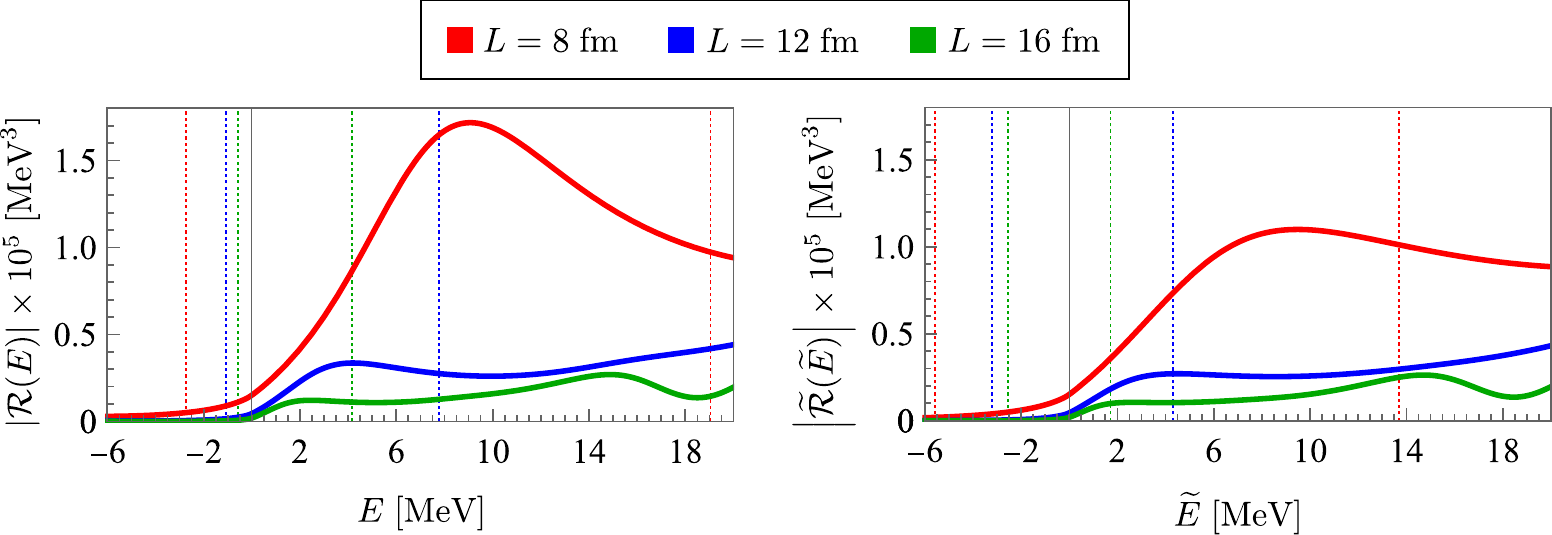}
    \caption{The absolute values of the LL residue function in the $^1S_0$ (left) and $^3S_1$ (right) channels is plotted against the CM energy for three different volumes with $L=8\;{\rm fm}$ (red), $L=12\;{\rm fm}$ (blue), and $L=16\;{\rm fm}$ (green). Dashed lines indicate energy eigenvalues in the respective volumes. The numerical values of $|\mathcal{R}|$ and $|\widetilde{\mathcal{R}}|$ evaluated at the FV ground- and first excited-state energies in the corresponding volumes are provided in Appendix \ref{app:detail}.
     \label{fig: R plot}}
\end{figure}

The cubic volume does not respect the rotational symmetry, and as a result, the FV quantization conditions mix scattering amplitudes in all partial waves. However, at low energies the scattering amplitude is expected to be dominated by the $S$-wave interaction. Ignoring the contribution from all higher-order partial waves, the FV quantization condition relates the $S$-wave phase shifts to a discrete set of FV energy eigenvalues, $E_n$. For NN systems in the $^1S_0$ channel, the quantization condition is given by
\begin{equation}
    p_n\cot{\delta} = 4\pi c_{00}(p_n^2,L).
    \label{eq: quantization condition}
\end{equation}
Here, $p_n=\sqrt{ME_n}$, and $\delta$ is the corresponding $S$-wave scattering phase shift. The FV function $c_{00}(p_n^2,L)$ is given by~\cite{Luscher:1986pf,Luscher:1990ux,Kim:2005gf} 
\begin{align}
c_{00}({p}_n^2,L)=\frac{1}{L\sqrt{4\pi^3}}\; \mathcal{Z}_{00}\left[1;(p_{n}L/2\pi)^2\right],~\text{with}~\mathcal{Z}_{00}[s;x^2]=\frac{1}{\sqrt{4\pi}}\sum_{\bm n \in \mathbb{Z}^3}\frac{1}{\left(|\bm{n}|^2-x^2\right)^s}.
\label{eq: Luscher's function}
\end{align}
Here, $\bm{n}$ is a Cartesian vector with integer components. The quantization condition in Eq.~\eqref{eq: quantization condition} is obtained by locating the singularities of the two-point correlation function of two nucleons, which is governed by the FV function $\mathcal{F}$,
\begin{align}
	\mathcal{F}(E) & = \frac{1}{F^{-1}_0(E)+\mathcal{M}(E)},
	\label{eq: definition of mathcal F}
\end{align}
where $F_0$ is another FV function related to the $c_{00}$ function defined above,
\begin{align}
    F_0(E)  =\frac{M}{4\pi}\left[-4\pi \; c_{00}({p}^2,L) +i p\right],
    \label{eq: F0 expression}
    \end{align}
and $\mathcal{M}$ is the NN scattering amplitude defined in Eq.~\eqref{eq: scattering amplitude}. Another useful quantity, which appears in the matching relations in Eqs.~\eqref{eq: matching relation for L1A} and~\eqref{eq: matching relation gvNN}, is the generalized Lellouch-L\"uscher (LL) residue matrix, $\mathcal{R}$, which is the residue of the FV function $\mathcal{F}$ at FV energies $E_n$, and is given by
\begin{align}
    \mathcal{R}(E_n) = \lim_{E \to E_n} (E-E_n)\;\mathcal{F}(E) = \bigg[\frac{d\mathcal{F}^{-1}}{dE}\biggr|_{E=E_n}\bigg]^{-1} .
    \label{eq: LL residue}
\end{align}
In the limit where higher partial waves are ignored, Eq.~\eqref{eq: quantization condition} is also valid for the $^3S_1$ channel after replacing $\delta$ with $\widetilde{\delta}$. Similarly, the replacement $\mathcal{M} \to \widetilde{\mathcal{M}}$ in Eq.~\eqref{eq: definition of mathcal F} defines $\mathcal{F}\to\widetilde{\mathcal{F}}$, which leads to the FV residue function $\widetilde{\mathcal{R}}$ for the $^3S_1$ channel.\par
\begin{figure}[t]
    \centering
    \includegraphics[scale=1]{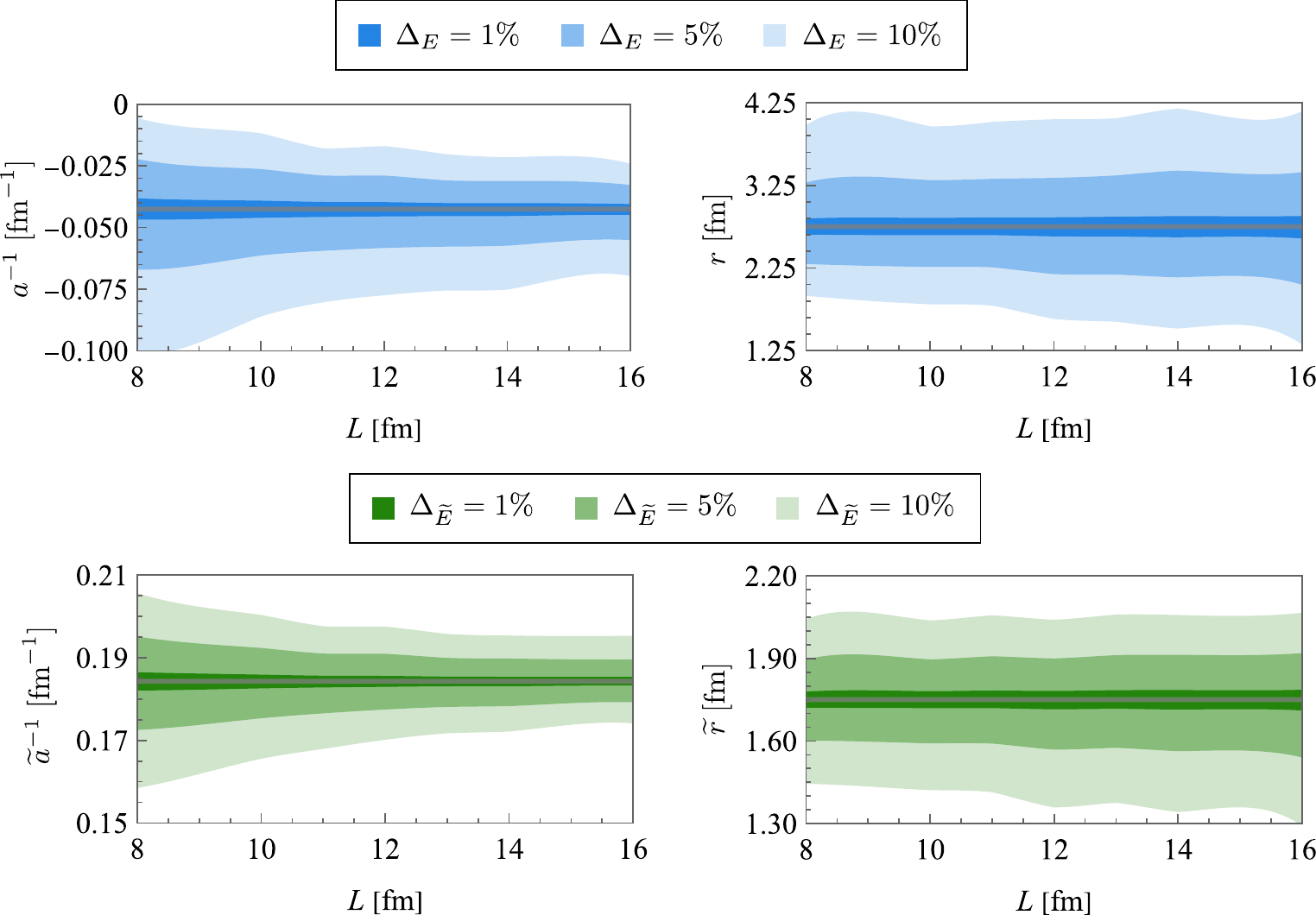}
    \caption{The inverse scattering length (left column) and the effective range (left column) for the $^1S_0$ (top row) and $^3S_1$ (bottom row) channels obtained from synthetic data with $\Delta_{E},\Delta_{\widetilde{E}}=10\%$, $5\%$, and $1\%$, from lighter to darker bands, respectively, are shown as a function of $L$. The bands indicate mid-$68\%$ uncertainty on the parameters from synthetic data, whereas gray thin bands denote the corresponding experimental values. Selected numerical values associated with this figure are provided in Appendix~\ref{app:detail}. 
    \label{fig: band ERE}}
\end{figure}
In the following sections, we investigate the sensitivity of constraining LECs $L_{1,A}$ and $g^{NN}_{\nu}$ to LQCD inputs from future LQCD calculations of the corresponding three- and four-point correlation functions at physical quark masses. The lowest-lying FV energy eigenvalues in each of the NN channels enter the necessary matching relations and these energies will be evaluated \emph{ab initio} from LQCD FV two-point correlation functions. As no LQCD determination of the FV spectrum at the physical quark masses exist to date, one can estimate the expected energies for given volumes by solving L\"uscher's quantization condition in Eq.~\eqref{eq: quantization condition} using experimental input for scattering amplitudes, as illustrated in Fig.~\ref{fig: QC plot}. Here, the function $p\cot \delta$ ($p\cot \widetilde{\delta}$) on the left-hand side of Eq.~\eqref{eq: quantization condition} is given by the effective-range expansion defined in Eq.~\eqref{eq: ERE} (and its counterpart for the $^3S_1$ channel). The effective-range expansion parameters
\begin{equation}
\begin{split}
    a & =-23.5\;[\text{fm}], \hspace{2cm} r = 2.75\;[\text{fm}],
    \\
    \widetilde{a} & =5.42\;[\text{fm}], \hspace{2.35cm} \widetilde{r} = 1.75\;[\text{fm}],
\end{split}
    \label{eq: ERE values}
\end{equation}
are obtained using NN phase shifts for $S$-wave scattering generated by the Nijmegen phenomenological NN potential~\cite{Stoks:1994wp}, that are the result of fits to NN scattering data in Ref.~\cite{NNonline}. The ground-state energies of the NN systems in the $^1S_0$ ($^3S_1$) channel with the CM energy $E_0$ ($\widetilde{E}_0$) for the volumes shown are negatively shifted compared with the threshold as noted in Fig.~\ref{fig: QC plot}, and asymptote polynomially (exponentially) to zero (to -2.2245 MeV) in the infinite-volume limit. Additionally, the absolute values of the LL residue functions are plotted in Fig.~\ref{fig: R plot} as a function of energy for $L=8,12$ and $16$~fm. Note that only the absolute values of these functions appear in the matching relations for the matrix elements.

The small uncertainties on the scattering parameters from experiment are ignored as the goal is to obtain central values of FV energies. For the sensitivity analyses of the upcoming sections, uncertainties need to be artificially introduced on these energies in generating synthetic data to mimic the expected LQCD uncertainties on energy extractions. This indicates that the scattering parameters associated with these energies will become uncertain too. Since the scattering parameters enter the LO and NLO NN scattering amplitudes, and hence impact the matching relations of the next sections, the subsequent uncertainty on scattering parameters must be taken into account. Uncertainties on the first two lowest-lying energies in each channel (which is a minimal set in a single volume to constrain the scattering length and effective range) can be introduced through a randomly-generated Gaussian distribution of energies with central values equal to $E_0$ and $E_1$ ($\widetilde{E}_0$ and $\widetilde{E}_1$) and the width equal to $\Delta_{E_0} \times |E_0|$ and $\Delta_{E_1} \times |E_1|$ ($\Delta_{\widetilde{E}_0} \times |\widetilde{E}_0|$ and $\Delta_{\widetilde{E}_1} \times |\widetilde{E}_1|$) for the ground- and first excited-state energies of the NN systems in the $^1S_0$ ($^3S_1$) channels, respectively. The scattering length and effective range corresponding to each channel for the choices of $\Delta_{E}\equiv\Delta_{E_0}=\Delta_{E_1}=10\%,~5\%$, and $1\%$ and $\Delta_{\widetilde{E}}\equiv\Delta_{\widetilde{E}_0}=\Delta_{\widetilde{E}_1}=10\%,~5\%$, and $1\%$ are then obtained by solving the quantization condition in Eq.~\eqref{eq: quantization condition}, resulting in uncertainties in the scattering parameters as shown in Fig.~\ref{fig: band ERE}. A similar analysis was performed in Ref.~\cite{Briceno:2013bda} in the isosinglet channel to study the viability of the extraction of the S-D mixing parameter from the upcoming LQCD calculations. 

Constraints on more than two energies, including in more than one volume and with various different boost vectors, will improve uncertainties on the extracted scattering parameters, possibilities that are not considered in this initial analysis. More radically, one may attempt to input the experimental determination of the scattering parameters (and hence the energy eigenvalues derived using quantization conditions) to avoid an uncertainty introduced in both quantities in costly LQCD calculations. This can reduce the uncertainty on the extracted LECs, as the only LQCD input will be matrix elements that are unknown experimentally. Nonetheless, the upcoming LQCD calculations will first evaluate these matrix elements at the isospin-symmetric limit where quantum electrodynamics (QED) effects and the non-vanishing mass difference among the light quarks are ignored. This means that for consistency, one needs to input the scattering parameters associated with the $^1S_0$ and $^3S_1$ channels in such a limit. As obtaining the isospin-symmetric parameters from experimental data involves model/EFT uncertainties, it is preferred that all inputs to the quantization and matching conditions are evaluated from first-principles LQCD calculations consistently. That is the strategy adopted in this synthetic data analysis. While the experimental parameters are used to obtain the central values of the FV energies, the subsequent analysis assumes energies and hence the scattering parameters are obtained directly from LQCD and hence involve likely sizable uncertainties in early calculations.\footnote{The inaccuracy in the central values of the FV energies compared to what is expected at the isospin symmetric limit will have minimal impact in the conclusions reached in the upcoming sections, as we have verified by slightly changing the central values of the synthetic data in our analysis and observed no significant sensitivity in achieved uncertainties on the LECs.}


\section{Sensitivity analysis for $L_{1,A}$
\label{sec: L1A}}
\noindent
At the NLO in the pionless EFT, the two-body axial-vector current contributes to single- and double-weak processes, including $pp$ fusion, neutrino(antineutrino)-induced disintegration of the deuteron, and muon capture on the deuteron~\cite{Butler:1999sv,Butler:2000zp,Davoudi:2020xdv}, and its strength is characterized by the LEC $L_{1,A}$. Constraints on $L_{1,A}$ were obtained using elastic and inelastic (anti)neutrino-deuteron scattering data from nuclear reactors: $L_{1,A}=3.6 \pm 5.5\text{ fm}^3$~\cite{Chen:1999tn}, as well as from Sudbury Neutrino Observatory~\cite{Bellerive:2016byv} and Super-K~\cite{Fukuda:2001nj,Fukuda:2002pe} experiments: $L_{1,A}=4.0 \pm 6.3\text{ fm}^3$~\cite{Butler:2002cw}. A more precise constraint was obtained in Ref.~\cite{Acharya:2019fij} where improved low-energy chiral EFT results of inelastic (anti)neutrino-deuteron scattering amplitude were matched to those of pionless EFT, resulting in: $L_{1,A}= 4.9 ^{+1.9}_{-1.5}\text{ fm}^3$. It is expected that the uncertainty in $L_{1,A}$ will be reduced to $\sim 1.25 \text{ fm}^3$ from the precise measurement of reaction rate of muon capture on the deuteron that is underway in the MuSun experiment~\cite{Andreev:2010wd}. 

Furthermore, a constraint on $L_{1,A}$ has been obtained from a LQCD study of the $pp$-fusion process in Ref.~\cite{Savage:2016kon} giving the value $L_{1,A}=3.9(0.2)(1.4) \text{ fm}^3$. Even though the statistical uncertainty shown in the first parentheses is small, the overall uncertainty is similar to the experimental constraints due to the large systematic uncertainty indicated in the second parentheses. The major source of uncertainty is the extrapolation to the physical quark masses as the correlation function for the $pp$-fusion process was calculated at larger quark masses corresponding to $m_\pi\approx 806$ MeV. Thus, it is expected that this uncertainty will improve in future LQCD calculations at lighter quark masses. However, the extraction of this LEC at such a large pion mass did not require the involved matching relation that will be presented shortly, as the NN states appeared deeply bound. Furthermore, achieving the quoted statistical uncertainty with quark masses near the physical values will be challenging. The question that will be addressed here is whether these features will limit the precise extraction of $L_{1,A}$ at the physical values of the quark masses.
\par

In this section, we investigate the accuracy with which $L_{1,A}$ can be obtained from future LQCD calculations performed at the physical pion mass. In Sec.~\ref{subsec: matching relation L1A}, the matching relation provided in Ref.~\cite{Davoudi:2020xdv} will be reviewed, relating the hadronic scattering amplitude for the $nn \to npe^-\bar{\nu}_e$ decay (or alternatively the $pp$ fusion process $pp \to npe^+\nu_e$) to the corresponding nuclear ME calculated using LQCD. The matching relation is then used to perform a sensitivity analysis on $L_{1,A}$ in Sec~\ref{subsec: sensitivity analysis L1A} through studying the effects of the LQCD inputs and their uncertainty on the $L_{1,A}$ extraction.

\subsection{Matching Relation
\label{subsec: matching relation L1A}}
Consider the single-$\beta$ decay transition $nn \to npe^-\bar{\nu}_e$ with the kinematics chosen such that the total three-momentum of the electron and anti-neutrino is zero, and the NN systems are unboosted in the initial and final states. The hadronic amplitude receives non-vanishing contribution from the Gamow-Teller-type transitions mediated by one-body ($n=1$) and two-body ($n=2$) axial-current operators, $A_{k(n)}^{i}$, where $k$ and $i$ denote spin and isospin indices, respectively. In the spin-isospin symmetric limit, the amplitude is independent of the azimuthal spin quantum number of the final state. Thus, one can consider the hadronic transition $nn \to np$ $(k=3)$ in the pionless EFT. The LO contribution is characterized by the corresponding LO NN contact interactions in Eq.~\eqref{eq: Nucleon EFT Lagrangian} for each channel and the one-body axial-vector current operator corresponding to the nucleon axial charge, $g_A$. At the NLO, the hadronic amplitude receives a contribution from the two-body axial-vector current operator corresponding to the LEC $L_{1,A}$:
\begin{equation}
	A_{3(2)}^{+}=L_{1,A}\big( N^{T}\widetilde{\mathcal{P}}_{3}N \big) ^{\dagger }\big( N^{T}\mathcal{P}_{+}N \big),
	\label{eq: isovector current : two body}
\end{equation}
where $\mathcal{P}_{+} = (\mathcal{P}_{1}+i \mathcal{P}_{2})/{\sqrt{2}}$.\par
The hadronic amplitude is related to the FV nuclear ME of the weak current between the ground states of the NN system with energies $E_0$ and $\widetilde{E}_0$ corresponding to the $^1S_0$ and $^3S_1$ channels, respectively, via the matching relation~\cite{Davoudi:2020xdv, Briceno:2012yi}
\begin{align}
    L^6&\left|\left[\vphantom{B^\dagger}\langle E_{0},L|\,\mathcal{J}({0})\,|\widetilde{E}_{0},L\rangle \right]_L\right|^2 = \left|\widetilde{\mathcal{R}}(\widetilde{E}_{0})\right| \left|\mathcal{M}^{\rm{DF},V}_{nn\to np} (E_{0},\widetilde{E}_{0})\right|^2  \left|\mathcal{R}(E_{0}) \vphantom{\widetilde{\mathcal{R}}(\widetilde{E}_{0})} \right|,
    \label{eq: matching relation for L1A}
\end{align}
where the equality is up to exponentially suppressed corrections in $L$. Here, $|\cdot|$ denotes the absolute value, and $\mathcal{J}$ denotes the hadronic part of the weak current placed at the origin, see Ref.~\cite{Davoudi:2020xdv}. The FV $S$-wave states, $|E,L\rangle$, are labeled with the CM energy, $E$, and the spatial extent of the cubic volume $L$, and the FV nature of the ME is emphasized by the subscript $L$. The quantity, $\mathcal{M}^{\rm{DF},V}_{nn\to np}$, is related to infinite-volume amplitude via
\begin{align}
   i \mathcal{M}^{\rm{DF},V}_{nn\to np} (E_0,\widetilde{E}_{0})=
   i\mathcal{M}^{\rm DF}_{nn\to np} & (E_0,\widetilde{E}_{0})-ig_A\,F_1(\widetilde{E}_{0},E_0) 
   \left[\mathcal{M}^{\rm (LO)}({E}_{0})\widetilde{\mathcal{M}}^{\rm (LO)}(\widetilde{E}_{0})+ \right.
   \nonumber
   \\
   & \left.\mathcal{M}^{\rm (LO)}({E}_{0})\widetilde{\mathcal{M}}^{\rm (NLO)}(\widetilde{E}_{0})+\mathcal{M}^{\rm (NLO)}({E}_{0})\widetilde{\mathcal{M}}^{\rm (LO)}(\widetilde{E}_{0})\right],
   \label{eq: L1A DF amplitude: finite V}
\end{align}
where $\mathcal{M}^{\rm DF}_{nn\to np}$ is the divergence-free infinite-volume amplitude, which is obtained after removing from the full amplitude the contributions from the Feynman diagrams with the weak current on the external nucleon legs. The LO and NLO NN scattering amplitudes in the $^1S_0$ channel, $\mathcal{M}^{\rm (LO)}$ and $\mathcal{M}^{\rm (NLO)}$, are defined in Eqs.~\eqref{eq: MLO NN scattering} and~\eqref{eq: MNLO NN scattering}, respectively. $\mathcal{R}$ is the LL residue function defined in Eq.~\eqref{eq: LL residue} for the $^1S_0$ channel. The corresponding quantities for the $^3S_1$ channel are denoted with an overhead tilde. $F_1$ is a FV function originating from the s-channel loop diagram with three nucleon propagators. It is related to the $F_0$ function defined in Eq.~\eqref{eq: F0 expression},
\begin{equation}
    F_1(\widetilde{E}_{0},E_0) =\frac{1}{E_0-\widetilde{E}_{0}}\,\left[F_0(\widetilde{E}_{0})-F_0(E_0)\right].
    \label{eq:F1def}
\end{equation}
Finally, the amplitude $\mathcal{M}^{\rm DF}_{nn\to np}$ depends on the LEC $L_{1,A}$,
\begin{align}
    i\mathcal{M}^{\rm DF}_{nn\to np} = -i\widetilde{L}_{1,A} \,\widetilde{\mathcal{M}}^{\rm LO}(\widetilde{E}_{0}) & \mathcal{M}^{\rm LO}(E_0)-ig_A\,I_1(\widetilde{E}_{0},E_0) 
   \left[\mathcal{M}^{\rm (LO)}({E}_{0})\widetilde{\mathcal{M}}^{\rm (LO)}(\widetilde{E}_{0}) \right.
   \nonumber\\
   & \left.+\mathcal{M}^{\rm (LO)}({E}_{0})\widetilde{\mathcal{M}}^{\rm (NLO)}(\widetilde{E}_{0})+\mathcal{M}^{\rm (NLO)}({E}_{0})\widetilde{\mathcal{M}}^{\rm (LO)}(\widetilde{E}_{0})\right],
    \label{eq: L1A  DF amplitude: inf volume}
\end{align}
where $\widetilde{L}_{1,A}$ is the renormalization-scale-independent combination of $L_{1,A}$ and the LO and NLO NN LECs introduced in Sec.~\ref{subec: Pionless EFT}:
\begin{equation}
    \widetilde{L}_{1,A} =
    \frac{L_{1,A}}{C_0\,\widetilde{C}_0}-\frac{g_A\,M}{2}\frac{(C_2+\widetilde{C}_2)}{C_0\,\widetilde{C}_0},
    \label{eq: L1A scale independent}
\end{equation}
and $I_1$ in is defined as
\begin{equation}
    I_1(\widetilde{E}_{0},E_0) =\frac{iM^{3/2}}{4\pi}\frac{1}{\sqrt{\widetilde{E}_0}+\sqrt{E_0 \vphantom{\widetilde{E}_0}} }.
    \label{eq: I1 definition}
\end{equation}
%

\subsection{Sensitivity analysis
\label{subsec: sensitivity analysis L1A}}
Future constraints on $L_{1,A}$ from LQCD calculations at the physical quark masses will depend on LQCD determinations of the low-lying FV energy eigenvalues of the NN systems in the $^1S_0$ and $^3S_1$ channels, as well as the nuclear MEs of the axial-vector current between these states, as is clear from the ingredients of Eq.~\eqref{eq: matching relation for L1A}. Furthermore, the matching relation depends upon the LO and NLO NN scattering amplitudes in both the $^1S_0$ and $^3S_1$ channels, as well as the derivative of scattering amplitudes with respect to energy that enters the LL residue function in Eq.~(\ref{eq: LL residue}), requiring the values of the scattering length and effective range in the two NN channels. These are obtained from the knowledge of at least two energy levels in the spectrum, i.e., the ground and the first excited states, as outlined in Sec.~\ref{subec: FV formalism}. The precision with which $L_{1,A}$ can be obtained depends on the precision and correlation of these ingredients. In order to quantify the uncertainty on $L_{1,A}$ extracted from a future LQCD calculation performed in a given volume, one can introduce percent precision with which the nuclear ME of a single axial-vector current and the NN ground- (and first excited-) state energies are expected to reach, to be denoted by $\Delta_{\beta}$ and $\Delta_{E(\widetilde{E})}$, respectively. A sample set of these ingredients is then generated from a Gaussian distribution with the mean represented by the value obtained from (the central values of) the phenomenological constraint for the quantity, and the  precision level multiplied by the mean for its standard deviation.

The mean values for the expected ground- and first excited-state energies of the NN channels are obtained using the quantization condition in Eq.~\eqref{eq: quantization condition} with the NN phase shifts for the $S$-channel from Ref.~\cite{NNonline}, as was already discussed in Sec.~\ref{subec: FV formalism} and demonstrated in Fig.~\ref{fig: QC plot}. The mean value of the expected FV ME is obtained by using the matching relation in Eq.~\eqref{eq: matching relation for L1A} with the FV energies being the mean values discussed above, the experimental value of $g_A\approx1.27$, and the central value of the $L_{1,A}$ (or and its scale-independent counterpart) from a recent phenomenological determination~\cite{Acharya:2019fij}
\begin{equation}
   L_{1,A}= 4.9 ^{+1.9}_{-1.5}\;\text{ fm}^3~~\text{or}~~\widetilde{L}_{1,A}= -449.7^{+19.5}_{-15.4}\;\text{ fm}^3.
    \label{eq: L1A value}
\end{equation}
\begin{figure}[t]
    \centering
    \includegraphics[scale=0.99]{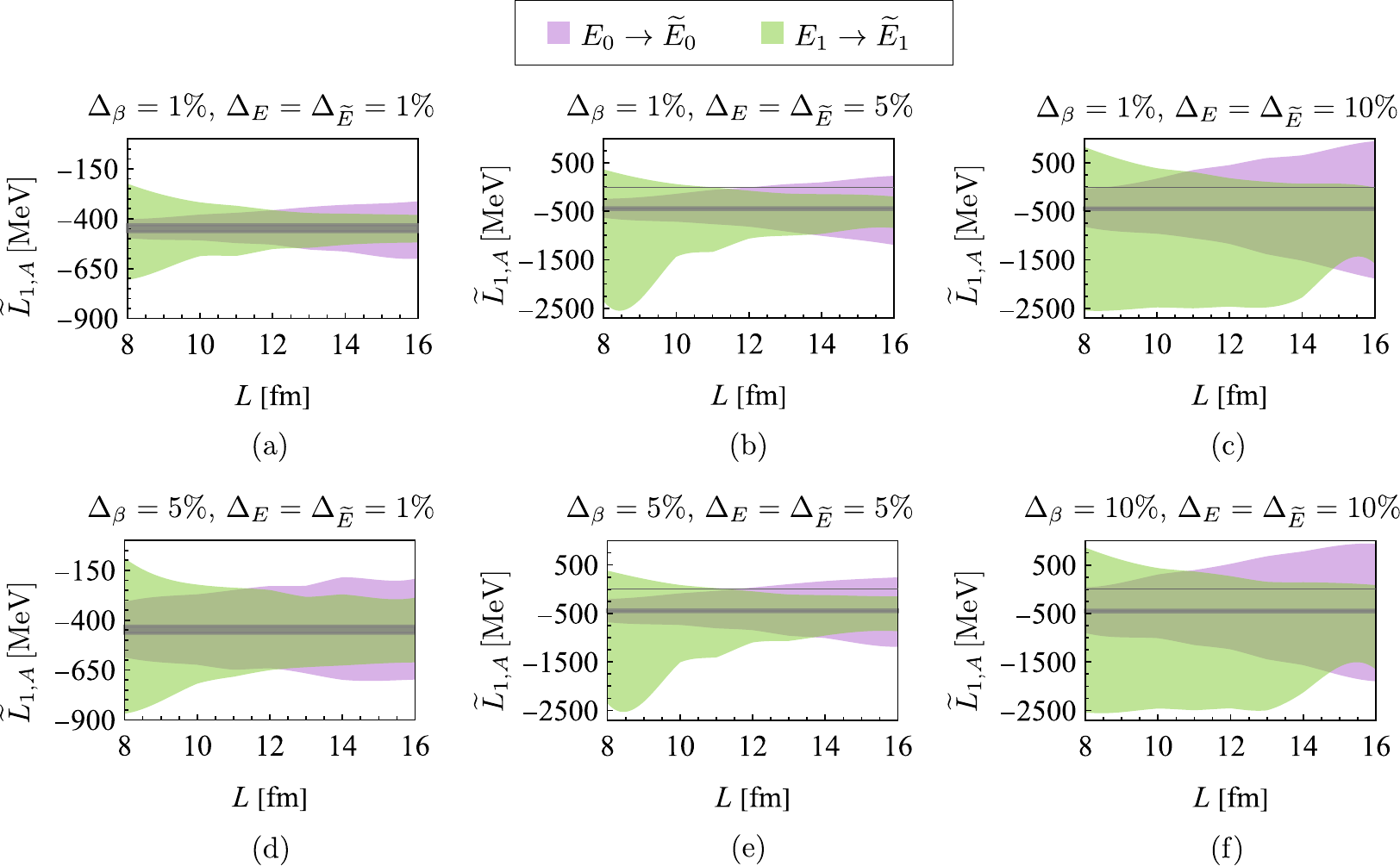}
    \caption{The value of $\widetilde{L}_{1,A}$ as a function of $L$ for the $^1S_0 \to {^3}S_1$ transition obtained from synthetic data with various combinations of $\Delta_{E}=\Delta_{\widetilde{E}}$ and $\Delta_{\beta}$ values. The gray horizontal band denotes the experimental value, whereas the colored bands indicate mid-$68\%$ uncertainty on extracted $L_{1,A}$ for the ground-state to ground-state  (purple) and first excited-state to first excited-state (green) transitions. Note the smaller range of the $\widetilde{L}_{1,A}$-axes in the most-left plots compared to the rest. Selected numerical values associated with this figure are provided in Appendix~\ref{app:detail}.
    \label{fig: band L1A}}
\end{figure}

The expected mean values are then used to generate the Gaussian samples for the FV energies and FV ME. With the samples generated, the matching relation in Eq.~\eqref{eq: matching relation for L1A} is used once again to solve for the $L_{1,A}$ values associated with each set of energies and MEs, leading to a distribution for the expected $L_{1,A}$ values. In the following, the scale-dependent quantity $\widetilde{L}_{1,A}$ is used but it can be converted to $L_{1,A}$ values give the values of the NN LECs evaluated at the corresponding values of the scattering length and effective range. Note that since the scattering parameters are obtained \emph{ab initio} from LQCD, the uncertainties in energies impact their precision, as discussed in Sec.~\ref{subec: FV formalism}.\par

The effect of $\Delta_{\beta}$ and $\Delta_E$ on determining $\widetilde{L}_{1,A}$ is illustrated in Fig.~\ref{fig: band L1A}, where the volume dependence of $L_{1,A}$ values obtained from the sample sets for various combinations of $\Delta_E$ and $\Delta_{\beta}$ values is shown. In all cases, the uncertainty on $L_{1,A}$ (determined from the mid-$68\%$ of the sample) increases with increasing $\Delta_E$, $\Delta_{\widetilde{E}}$, and $\Delta_\beta$. Only the most precisely determined sample set and at volumes with $L \approx 8\;{\rm fm}$, constraints on $L_{1,A}$ become comparable in precision to that in Eq.~\eqref{eq: L1A value}. Thus, future LQCD calculations at the physical quark masses need to determine the NN ground and first excited-state energies and the FV MEs with below percent-level precision to supersede the current phenomenological constraints. The situation is likely alleviated in the actual LQCD calculations where energy and ME extractions are partially correlated, and where the NN scattering amplitude can be determined more precisely with a larger set of precise FV energies.

Since LQCD can, in principle, obtain FV MEs for transitions involving excited states, one may wonder if constraining $L_{1,A}$ through the first excited-state to the first excited-state transition will be more beneficial and relaxes the precision requirements on the FV energies and ME above. The green bands in Fig.~\ref{fig: band L1A} denote the $\widetilde{L}_{1,A}$ values and uncertainties obtained from the first excited-state to the first excited-state transition. It is clear that the ground-state to ground-state transition leads to better constraints at smaller volumes---volumes that are more readily accessibly to upcoming LQCD calculations at the physical pion mass, but for larger volumes with $L \gtrsim 14\;{\rm fm}$, the constraints from the first excited-state to the first excited-state transition become comparable or more precise. The reverse trend in uncertainties as a function of volume between the two cases is a consequence of different behavior of the LL residue functions near negative and positive CM energies, as illustrated in Fig.~\ref{fig: R plot}. One cautionary note is the loss of accuracy in using the effective-range expansion and the associated LO and NLO NN scattering amplitudes in the pionless EFT near the first excited-state energies. However, at  large volume where the $\widetilde{L}_{1,A}$ constraints from excited-state transition become more precise, the FV energies tend to their asymptotic value of zero and are therefore near or within the t-channel cut. On the other hand, at such large volumes, the density of states in the spectrum increases, and the identification of excited states with current methods may present a challenge. Variational techniques such as those developed in Refs.~\cite{Horz:2020zvv, Green:2021qol, Amarasinghe:2021lqa} will likely constrain the lowest-lying levels with comparable precision to the ground state.


\section{Sensitivity analysis for $g_{\nu}^{NN}$
\label{sec: gvNN}}
\noindent
In the light neutrino exchange model of the low-energy $nn \to pp e^-e^-$ decay, there exists an undetermined LEC, $g_{\nu}^{NN}$, at the LO the pionless EFT, which is introduced to absorb the UV scale dependence of the amplitude through renormalization group~\cite{Cirigliano:2017tvr,Cirigliano:2018hja,Cirigliano:2019vdj}. The Lagrangian density corresponding to this short-distance contribution consists of a four-nucleon-two-electron contact interaction:
\begin{equation}
    \mathcal{L}_{N}^{\Delta L =2} = \left(\frac{4V_{ud}G_{F}}{2\sqrt{2}}\right)^2 \,  m_{\beta\beta } \, g_\nu^{NN} \left[\overline{e}_L\, C\, \bar{e}^T_L \right] \left[ (N^T\mathcal{P}_-N)^\dagger N^T\mathcal{P}_+N)\right] + {\rm H.c.}
    \label{eq: gvNN operator}
\end{equation}
Here, $G_f$ is Fermi's constant, $V_{ud}$ is a Cabibbo-Kobayashi-Maskawa (CKM) matrix element~\cite{Cabibbo:1963yz, Kobayashi:1973fv}, $m_{\beta\beta }$ is the effective Majorana mass, $m_{\beta\beta} = \sum_i U^2_{ei} m_i$, where $U_{ei}$ are the elements of the Pontecorvo-Mako-Nakagawa-Sato (PMNS) matrix~\cite{Pontecorvo:1957qd, Maki:1962mu}, with $m_i$ being the mass of the neutrino-mass eigenstate $i$. $C$ is the charge-conjugation matrix, and $e_L$ is a left-handed electron field.\par

As shown in Refs.~\cite{Cirigliano:2020dmx,Cirigliano:2021qko}, a constraint on the $g_{\nu}^{NN}$ value can be obtained by expressing the  $nn \to pp e^-e^-$ decay amplitude as a product of momentum integral of the Majorana neutrino propagator and the generalized forward Compton scattering amplitude, in analogy to the
Cottingham formula~\cite{Cottingham:1963zz,PhysRevLett.17.1303} for the electromagnetic contribution to hadron masses. A model-independent representation of the integrand using the chiral EFT and operator product expansion can then be obtained. The missing parts of the full amplitude can be filled by interpolating between the known regions using nucleon form factors for the weak current and information on NN scattering. The constraint on $g_{\nu}^{NN}$ via this method is:
\begin{equation}
    \widetilde{g}_{\nu}^{NN} = 1.3 \pm 0.6,
    \label{eq: gvNNtilde value}
\end{equation}
where $\widetilde{g}_{\nu}^{NN}$ is a dimensionless parameter related to $g_{\nu}^{NN}$ and the momentum-independent NN LEC in Eq.~\eqref{eq: Nucleon EFT Lagrangian}: 
\begin{equation}
\widetilde{g}_{\nu}^{NN}= \bigg(\frac{4\pi}{M C_0}\bigg)^2 g_{\nu}^{NN}\,.
\label{eq: gvNNtilde}
\end{equation}
The value in Eq.~\eqref{eq: gvNNtilde value} has a large uncertainty, and a more precise and direct constraint on  $g_{\nu}^{NN}$ using LQCD will be desired. As shown in Ref.~\cite{Davoudi:2020gxs}, a prescription exists for obtaining the $g_{\nu}^{NN}$ (or equivalently the $\widetilde{g}_{\nu}^{NN}$) value from a Euclidean four-point correlation function calculated using LQCD. With LQCD calculations of these correlation functions underway, it would be useful to know the precision with which one can constrain the  $\widetilde{g}_{\nu}^{NN}$ value for a given LQCD setup. In this section, we perform the sensitivity analysis of constraining  $\widetilde{g}_{\nu}^{NN}$ by estimating the uncertainty on  $\widetilde{g}_{\nu}^{NN}$ from a synthetic data representing a future LQCD calculation of the four-point correlation function at the physical quark masses.\par

\subsection{Matching Relation
\label{subsec: matching relation gvNN}}
Consider the transition $nn \to pp e^-e^-$ in the spin-isospin symmetric limit with simple kinematics, where the currents carry zero energy and momentum such that the initial CM energy, $E_i \equiv E_{CM}$, remains unchanged. The Euclidean four-point function for this process, which is accessible via LQCD methods, can be analytically continued to Minkowski spacetime to obtain
\begin{eqnarray}
\mathcal{T}^{(\rm M)}_L \equiv \int dz_0\,\int_L d^3z \,
\big[\langle E_0,L|\, T[\mathcal{J}(z_0,\bm{z})\,S_\nu(z_0,\bm{z})
\mathcal{J}(0)]\, |E_0,L\rangle\big]_L,
\label{eq: TML definition}
\end{eqnarray}
using the procedure described in Ref.~\cite{Davoudi:2020gxs}. In Eq.~\eqref{eq: TML definition}, $T$ denotes time ordering, the superscript $(\rm M)$ denotes a Minkowski time signature, the subscript $L$ on the spatial integral indicates that the integral is performed over a finite cubic volume (with PBCs), and $z_0$ is the Minkowski time coordinate. $S_\nu$ is the Minkowski propagator of a Majorana neutrino in a finite volume that is given by
\begin{equation}
S_\nu(z_0,\bm{z}) = \frac{1}{L^3}\sum_{\bm{k}  \in \frac{2\pi}{L}\mathbb{Z}^3\neq \bm{0}} \int \frac{dk_0}{2\pi}e^{i\bm{k} \cdot \bm{z}-ik_0z_0} \frac{-i\,m_{\beta\beta}}{k_0^2-|\bm{k}|^2+i\epsilon},
\label{eq: Neutrino propagator Minkowski}
\end{equation}
where the neutrino four-momentum is given by $(k_0,\bm{k})$ with quantized spatial momenta $\bm{k}$. Contributions from the small non-zero neutrino mass in the denominator of the neutrino propagator can be ignored at the LO in the EFT power counting, and the infrared divergence is regulated by removing the zero-momentum mode of the neutrino. The remaining notation in Eq.~\eqref{eq: TML definition} is the same as in Eq.~\eqref{eq: matching relation for L1A}. It is important to note that for a ground-state to ground-state transitions at low energies corresponding to the FV energy eigenvalues in the range of volumes studied, no intermediate single-neutrino-two-nucleon state can go on shell and the analytic continuation from the Euclidean correlation function of LQCD to the Minkowski counterpart in Eq.~(\ref{eq: TML definition}) is straightforward. With on-shell intermediate states, the complete formalism of Ref.~\cite{Davoudi:2020gxs} needs to be implemented but this will not be necessary in the upcoming LQCD calculations given realistic volumes and energies.

$\mathcal{T}^{(\rm M)}_L$ is related to the physical decay amplitude through the following matching relation:
\begin{equation}
L^6\;\bigg|\mathcal{T}^{(\rm M)}_L(E_i,E_f) \bigg|^2=\bigg|\mathcal{R}(E_i)\bigg | \, \bigg|\mathcal{M}^{0\nu,V}_{nn\to pp} (E_i,E_f) \bigg|^2 \bigg|\mathcal{R}(E_f)\bigg |,
\label{eq: matching relation gvNN}
\end{equation}
where
\begin{equation}
    \mathcal{M}^{0\nu,V}_{nn\to pp} (E_{i},E_{f}) = \mathcal{M}^{(\rm{Int.})}_{nn\to pp} (E_{i},E_{f})-m_{\beta\beta}(1+3g_A^2)\mathcal{M}^{(\rm LO)}(E_i)\delta J^V(E_{i},E_{f}) \mathcal{M}^{(\rm LO)}(E_f).
    \label{eq: Finite volume amplitude 0vbb}
\end{equation}
The right-hand side of Eq.~\eqref{eq: matching relation gvNN} contains the LL residue matrix, $\mathcal{R}$, defined in Eq.~\eqref{eq: LL residue}, and the FV quantity $\mathcal{M}^{0\nu,V}_{nn\to pp}$ which is related to the physical scattering amplitude of the $0\nu\beta\beta$ decay with the initial (final) CM energy $E_i$ $(E_f)$, as defined in Eq.~\eqref{eq: Finite volume amplitude 0vbb}. Here, $\mathcal{M}^{({\rm LO})}$ is the LO NN scattering amplitude defined in Eq.~\eqref{eq: MLO NN scattering}, $\mathcal{M}^{(\rm{Int.})}$ is the infinite-volume decay amplitude evaluated in the pionless EFT after removing the contributions from the diagrams in which the neutrino propagates between two external nucleons. The full scattering amplitude is evaluated assuming that the amplitude is approximated by the s-wave interactions of the nucleons and only receives contributions from a static neutrino potential. Moreover, contributions to the full infinite-volume amplitude from radiative neutrinos are ignored. With these assumptions, $\mathcal{M}^{(\rm{Int.})}$ is given by~\cite{Cirigliano:2017tvr,Cirigliano:2018hja,Cirigliano:2019vdj} 
\begin{eqnarray}
\mathcal{M}^{(\rm{Int.})}_{nn\to pp} (E_i,E_f)=m_{\beta\beta}\;\mathcal{M}^{(\rm LO)}(E_i) \bigg [ -(1+3g_A^2) J^{\infty}(E_i,E_f;\mu)+\frac{2g_\nu^{NN}}{C_0^2}\bigg]
\mathcal{M}^{(\rm LO)}(E_f).
\label{eq: M^int}
\end{eqnarray}
The first term denotes contributions from the diagrams in which the neutrino propagates between two nucleons dressed by strong interactions on both sides. $J^\infty$ is a known function given by
\begin{equation}
J^{\infty}(E_i,E_f;\mu)=\frac{M^2}{32\pi^2} \bigg[-\gamma_E+\ln(4\pi)+\ln\left(\tfrac{\mu^2/M}{-(\sqrt{E_i}+\sqrt{E_f})^2-i\epsilon}\right)+1\bigg],
\label{eq: Jinf}
\end{equation}
with $\gamma_E$ being Euler's constant. This arises from evaluating the s-channel two-loop diagram with an exchanged Majorana neutrino. The UV divergence is regularized in the dimensional-regularization scheme, introducing the scale $\mu$. The second term in square brackets Eq.~\eqref{eq: M^int} denotes contributions from diagrams with the NN short-range operator in Eq.~\eqref{eq: gvNN operator} dressed by the NN propagator and the LO NN amplitude on both sides. Finally, $\delta J^V(E_i,E_f)$ in Eq.~\eqref{eq: matching relation gvNN} is a FV function corresponding to the FV two-loop diagram with the exchanged neutrino propagator, that is defined by
\begin{eqnarray}
\delta J^V(E_i,E_f)=
\bigg[\frac{1}{{L^6}}\sum_{\substack{\bm{k}_1,\bm{k}_2  \in \frac{2\pi}{L}\mathbb{Z}^3\\ \bm{k}_1 \neq \bm{k}_2}}-
\int \frac{d^3k_1}{(2\pi)^3}\frac{d^3k_2}{(2\pi)^3}
\bigg]
\frac{1}{E_i-\tfrac{|\bm{k}_1^2|}{M}+i\epsilon} \frac{1}{E_f-\tfrac{|\bm{k}_2^2|}{M}+i\epsilon}\frac{1}{|\bm{k}_1-\bm{k}_2|^2}.
\label{eq: deltaJV}
\end{eqnarray}
\begin{figure}[t]
    \centering
    \includegraphics[scale=1]{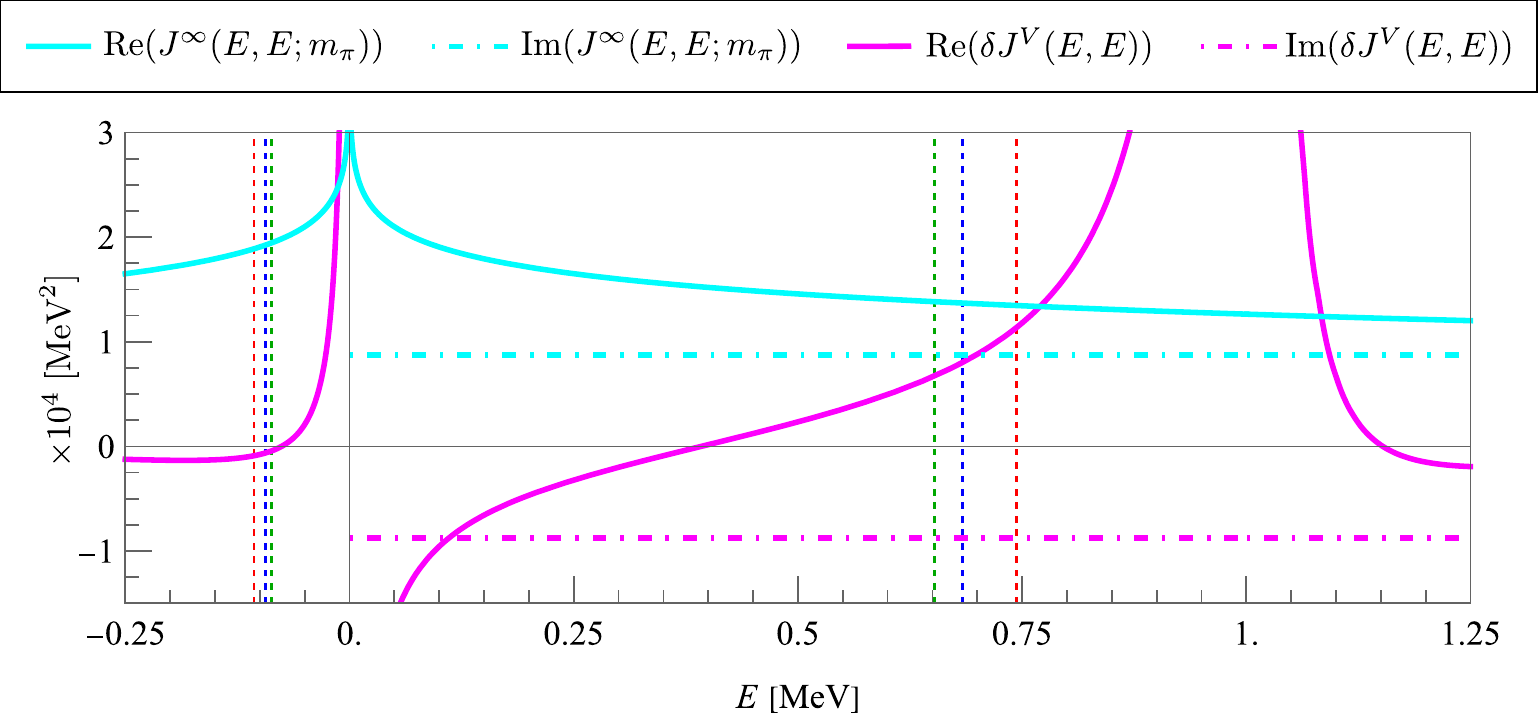}
    \caption{The real (solid cyan) and imaginary (dotted dashed cyan) parts of $J^{\infty}(E_i,E_f;\mu=m_\pi)$ defined in Eq.~(\ref{eq: Jinf}), as well as real (solid magenta) and imaginary (dotted dashed magenta) parts of $\delta J^V(E_i,E_f)$ defined in Eq.~(\ref{eq: deltaJV}), both evaluated at $E_i=E_f\equiv E$. The red, blue, and green dashed lines denote the FV ground-state energy eigenvalues with $L=8$, $12$, and $16\;{\rm fm}$, respectively, obtained from the quantization condition in Eq.~(\ref{eq: quantization condition}) (as plotted in Fig.~\ref{fig: QC plot}). These are the values at which the LQCD four-point function will be evaluated in the future studies at the physical quark masses.  Selected numerical values for the functions shown are provided in Appendix~\ref{app:detail}.
    \label{fig:delJ}
    }
\end{figure}
\begin{figure}[t]
    \centering
    \includegraphics[scale=1]{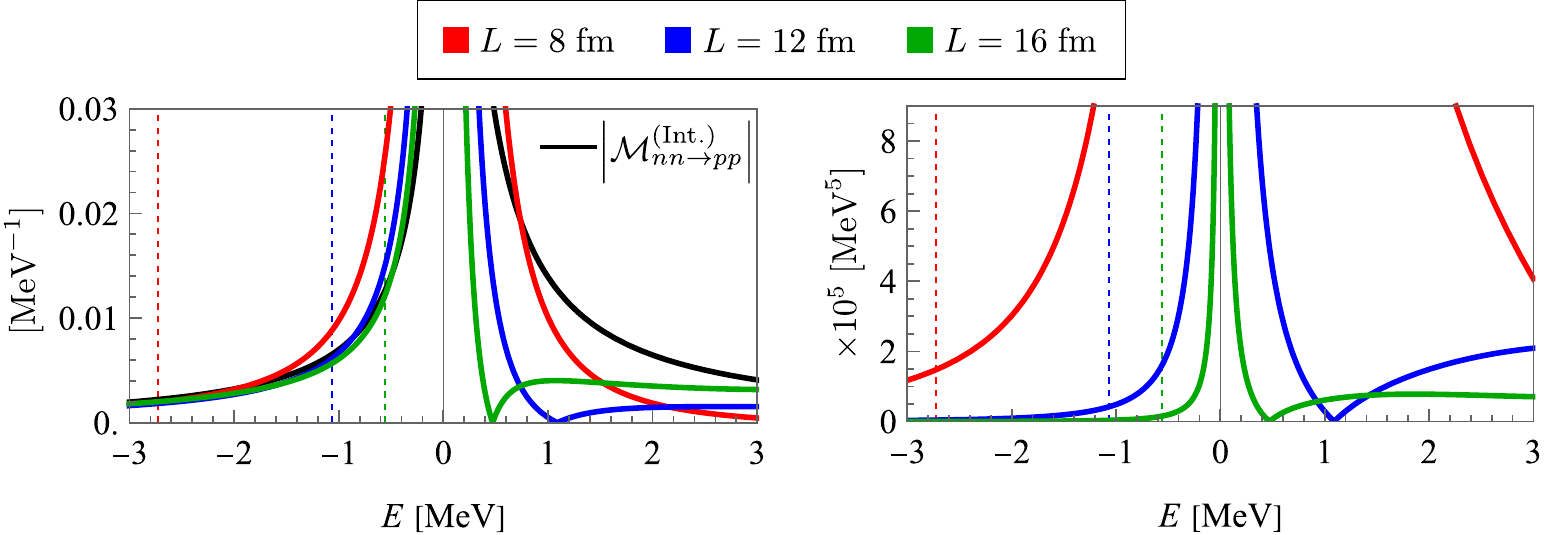}
    \caption{$\big| \mathcal{M}^{0\nu,V({\rm Int.})}_{nn\to pp}\big|$ (left) and $\big|\mathcal{T}_L^{(M)}\big|$ (right) functions defined in Eqs.~(\ref{eq: M^int})-(\ref{eq: matching relation gvNN}), with $L=8\;{\rm fm}$ (red), $L=12\;{\rm fm}$ (blue), and $L=16\;{\rm fm}$ (green) are plotted against the CM energy of the NN state, considering the kinematics $E_i=E_f \equiv E$. The effective neutrino mass $m_{\beta\beta}$ is set to $1\;{\rm MeV}$. The dashed lines in both panels denote the ground-state energy eigenvalues in the corresponding volumes obtained from the quantization condition in Eq.~(\ref{eq: quantization condition}) (as plotted in Fig.~\ref{fig: QC plot}). Selected numerical values for the functions shown are provided in Appendix~\ref{app:detail}.
    \label{fig: TLMMV vs ECM}
    }
\end{figure}
This sum-integral difference is calculated numerically using the technique presented in the supplemental material of Ref.~\cite{Davoudi:2020gxs}. The real and imaginary parts of $J^{\infty}$ and $\delta J^V$ are depicted in Fig.~\ref{fig:delJ} for a range of negative and positive $E_i=E_f\equiv E$ values.

The absolute value of the FV amplitude $\mathcal{M}^{0\nu,V}_{nn\to pp}$ for the kinematics $E_i=E_f \equiv E$ is plotted against the CM energy in the left panel of Fig.~\ref{fig: TLMMV vs ECM} along with $|\mathcal{M}^{\rm (Int.)}_{nn \to pp}|$, using the value of $g_{\nu}^{NN}$ obtained from the central value of the constraint in Eq.~\eqref{eq: gvNNtilde value}. The dependence of the $|\mathcal{T}_L^{(M)}|$ on the CM energy of the NN system in different volumes is shown in the right panel of Fig.~\ref{fig: TLMMV vs ECM}(b) using the matching relation in Eq.~\eqref{eq: matching relation gvNN}. 

\subsection{Sensitivity Analysis
\label{subsec: sensitivity analysis gvNN}}
Equation~\eqref{eq: matching relation gvNN} indicates that the precision with which $g_{\nu}^{NN}$, and thus $\widetilde{g}_{\nu}^{NN}$, can be obtained from LQCD depends on the precision with which the FV ground-state energy in a given volume, $E_0$, and the FV ME are obtained from the LQCD calculations of the corresponding two- and four-point functions, respectively. Furthermore, the matching relation depends upon the LO NN scattering amplitude in the $^1S_0$ channel as well as the derivative of the NLO+LO scattering amplitude with respect to energy that enters the LL residue function in Eq.~(\ref{eq: LL residue}), requiring the values of the scattering length and effective range in the $^1S_0$ channel. These depend on the central value and the uncertainty of at least two energy levels in the spectrum, e.g., the ground and the first excited states, as outlined in Sec.~\ref{subec: FV formalism}. In this section, we investigate the uncertainty on $\widetilde{g}_{\nu}^{NN}$ from the precision levels with which these LQCD inputs are obtained in future LQCD calculations at the physical quark masses.

The expected value of $E_0$ for a given volume is calculated using L\"uscher's quantization condition in Eq.~\eqref{eq: quantization condition} and NN phase shifts in the $^1S_0$ channel obtained from Ref.~\cite{NNonline}. This expected value of $E_0$ and the central value of the constraint on $\widetilde{g}_{\nu}^{NN}$ given in Eq.~\eqref{eq: gvNNtilde value} are then used to obtain an estimate on the expected value of $\mathcal{T}^{(\rm M)}_L$ with the use of Eqs.~\eqref{eq: matching relation gvNN} and~\eqref{eq: M^int}. Note that even though the expected value of $\mathcal{T}^{(\rm M)}_L$ from Eqs.~\eqref{eq: TML definition}-\eqref{eq: M^int} is dependent on $m_{\beta\beta}$, the mean value and the uncertainty on $\widetilde{g}_{\nu}^{NN}$ obtained from synthetic data using Eq.~\eqref{eq: matching relation gvNN} is independent of $m_{\beta\beta}$.

\begin{figure}[t]
    \centering
    \includegraphics[scale=1]{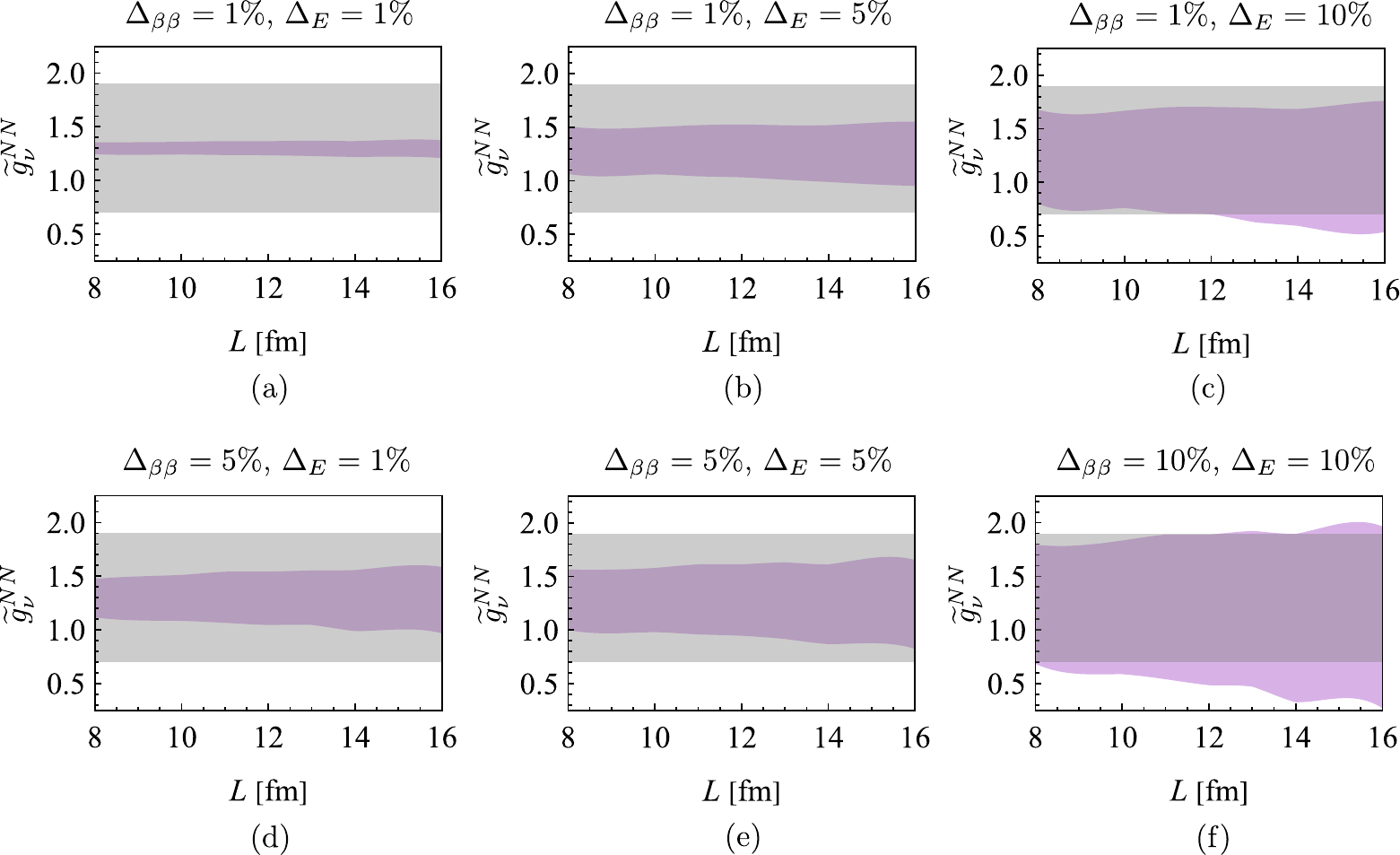}
    \caption{ The value of $\widetilde{g}_{\nu}^{NN}$ obtained from the synthetic data is plotted against $L$ for different combinations of $\Delta_{\beta\beta}$ and $\Delta_E$. The gray band denotes the uncertainty in the value of $\widetilde{g}_{\nu}^{NN}$ from Eq.~\eqref{eq: gvNNtilde value} from the indirect determination of Ref.~\cite{Cirigliano:2020dmx}. The corresponding central value is used to obtain the expected values of $\mathcal{T}^{(\rm M)}_L$, which enables this sensitivity analysis.  The purple band is the mid-$68\%$ uncertainty band corresponding to the sample sets with uncorrelated fluctuations. Selected numerical values associated with this figure are provided in Appendix~\ref{app:detail}.
    \label{fig: band gvNN}}
\end{figure}
The percent precision on $E_0$ (and $E_1$) is denoted by $\Delta_E$, whereas the percent precision on $\mathcal{T}^{(\rm M)}_L$ is denoted by $\Delta_{\beta\beta}$. Similar to Sec.~\ref{subsec: sensitivity analysis L1A}, the uncertainty on $\widetilde{g}_{\nu}^{NN}$ is taken as the mid-$68\%$ of the ensemble of $\widetilde{g}_{\nu}^{NN}$ values obtained from synthetic data that incorporates uncertainties on $E_{0(1)}$ and $\mathcal{T}^{(\rm M)}_L$ as Gaussian fluctuations. The precision levels, $\Delta_E$ and $\Delta_{\beta\beta}$, are incorporated in this synthetic data by making the standard deviation of the fluctuations equal to the expected values of the quantities multiplied by the corresponding percent precision. The scattering length and effective range in the $^1S_0$ channel are obtained by solving L\"uscher's quantization condition in Eq.~\eqref{eq: quantization condition} for the generated ensembles of the ground- and the first excited-state energies, as outlined in sec.~\ref{subec: FV formalism}.

The $\widetilde{g}_{\nu}^{NN}$ values obtained for various combinations of $\Delta_{\beta\beta}$ and $\Delta_E$ are plotted against $L$ in Fig.~\ref{fig: band gvNN}. The LQCD constraints on $\widetilde{g}_{\nu}^{NN}$ are almost always more precise than the constraint of Ref.~\cite{Cirigliano:2020dmx} for input uncertainties below $\sim10\%$ level, which indicates that future LQCD calculations can confidently improve the current constraint, especially for smaller volumes, provided that $\Delta_{\beta\beta}$ and $\Delta_E$ are a few percents. This situation is more promising than the case of $L_{1,A}$, where (sub)precent-level uncertainties appear to be the requirement. As the LQCD input for energies and the ME will be partially correlated, the constraint on $\widetilde{g}_{\nu}^{NN}$ will likely be further improved.


\section{Conclusions
\label{sec: conclusions}}
\noindent
This paper presents an analysis of the effect of uncertainties in the future lattice quantum chromodynamics calculations at the physical quark masses on the accuracy with which the hadronic amplitudes of $\beta$ decays can be constrained in the two-nucleon sector. The nuclear matrix elements of the single-$\beta$ decay and the neutrinoless double-$\beta$ decay within the light neutrino exchange scenario are studied for this purpose, and the precision with which the low-energy constants $L_{1,A}$ and $\tilde{g}_{\nu}^{NN}$, corresponding to the respective two-body isovector and isotensor operators, can be obtained from future calculations was deduced from a synthetic data analysis.
\par

For processes that are studied here, matching relations exist that relate the three- and four-point functions of LQCD evaluated in a finite Euclidean spacetime to their respective physical scattering amplitudes~\cite{Briceno:2012yi,Detmold:2004qn,Briceno:2015tza,Davoudi:2020xdv,Davoudi:2020gxs}. The LQCD inputs that go into these matching relations involve the lowest-lying two-nucleon energy spectra for a given volume, the matrix elements of a single axial-vector weak current (for the single-$\beta$ decay), and of two axial-vector weak currents along with a Majorana neutrino propagator (for the $0\nu\beta\beta$ decay) between appropriate two-nucleon states. Using these matching relations, constraints were obtained on $L_{1,A}$ and $\tilde{g}_{\nu}^{NN}$ from the synthetic data of the relevant LQCD ingredients. In order to synthesize this data to represent the underlying LQCD uncertainties, Gaussian fluctuations were introduced on the supposedly LQCD ingredients that go into these matching relations. 

The precision with which $L_{1,A}$ and $\tilde{g}_{\nu}^{NN}$ can be obtained from the synthetic data was obtained for a range of input uncertainties at or below $\sim10\%$ level. The uncertainty on the LECs grows with volume in both cases assuming ground-state to ground-state transitions, and so smaller volumes that are more feasible computationally appear to be more advantageous. The constraints from LQCD studies on $L_{1,A}$ will likely be worse than the current experimental constraints for the range of volumes and plausible input uncertainties considered here, and may require (sub)percent-level precision on the finite-volume energies and matrix element. The situation may be alleviated in actual LQCD calculations where the uncertainties in the inputs to the matching relations are (partially) correlated. Furthermore, one may imagine inputting the precise experimental parameters and associated FV energies in those analyses, rather than obtaining them directly from LQCD calculations, to decrease the uncertainty in the extraction of the unknown LECs. Nonetheless, such an approach will not be \emph{ab initio}, particularly since the early calculations will take place at the isospin-symmetric limit and excluding QED, and for consistency and model independency, scattering parameters need to be evaluated directly from LQCD. 

Finally, for precision levels on the LQCD energies and the ME below $10\%$, the constraint on $\tilde{g}_{\nu}^{NN}$ will likely improve the existing constraint, and will therefore provide a direct precise determination arising from first-principles calculations rooted in QCD. As a result, the present study further motivates future studies of the $nn \to ppee$ process within the light Majorana exchange scenario from LQCD at or near the physical values of the quark masses.

\section*{Acknowledgment}
\noindent
We acknowledge valuable discussions on a range of topics at the Institute for Nuclear Theory's virtual program on ``Nuclear Forces for Precision Nuclear Physics'' (INT-21-1b held in Spring 2021) which inspired the need for the analysis of this work. ZD and SVK are supported by the Alfred P. Sloan fellowship and by the Maryland Center for Fundamental Physics at the University of Maryland, College Park.


\appendix
\section{Numerical values associated with the figures
\label{app:detail}}
\noindent
Tables~\ref{tab: E and R table}-\ref{tab: gvNN} below contain many representative numerical values associated with the plots throughout the main text.
\begin{table}[h!]
    \renewcommand{\arraystretch}{1.5}
    \centering
    \begin{tabular}{C{1cm}|C{1.25cm}|C{1.75cm}|C{1.25cm}|C{1.75cm}|C{1.25cm}|C{1.75cm}|C{1.25cm}|C{1.75cm}}
        \hline
        \hline
        \rule{0pt}{4ex}
         $L$ & $E_0$ & $\left| \mathcal{R}(E_0) \right|$ & $E_1$ & $\left| \mathcal{R}(E_1) \right|$ &  $\widetilde{E}_0$ & $\left| \widetilde{\mathcal{R}}(\widetilde{E}_0) \right|$ & $\widetilde{E}_1$ & $\left| \widetilde{\mathcal{R}}(\widetilde{E}_1) \right|$ \\
         $\left[\text{fm}\right]$ &  $\left[\text{MeV}\right]$ &  $\left[\text{MeV}^3\right]$ &   $\left[\text{MeV}\right]$ & $\left[\text{MeV}^3\right]$ & $\left[\text{MeV}\right]$ & $\left[\text{MeV}^3\right]$ &  $\left[\text{MeV}\right]$ & $\left[\text{MeV}^3\right]$ \\
         \hline
         8 & -2.728 & $5.04\times 10^3$ & 19.043 & $9.73\times 10^4$ & -5.579 & $1.88\times 10^3$ & 13.688 & $1.01\times 10^5$ \\
         10 & -1.618 & $2.62\times 10^3$ & 11.606 & $4.87\times 10^4$ & -4.004 & $6.90\times 10^2$ & 7.364 & $4.94\times 10^4$ \\
         12 & -1.067 & $1.55\times 10^3$ & 7.772 & $2.75\times 10^4$ & -3.218 & $2.65\times 10^2$ & 4.299 & $2.70\times 10^4$ \\
         14 & -0.752 & $9.93\times 10^2$ & 5.560 & $1.69\times 10^4$ & -2.788 & $1.02\times 10^2$ & 2.655 & $1.59\times 10^4$ \\
         16 & -0.556 & $6.78\times 10^2$ & 4.176 & $1.11\times 10^4$ & -2.544 & $3.77\times 10^1$ & 1.712 & $9.85\times 10^3$ \\
         \hline
         \hline
    \end{tabular}
    \caption{Numerical values of the FV ground- and first excited-state energies in the $^1S_0$ and $^3S_1$ channels for a range of $L$ values along with the absolute values of the corresponding LL residue functions evaluated at those energies.}
    \label{tab: E and R table}
\end{table}
\begin{table}[hbt!]
    \renewcommand{\arraystretch}{1.75}
    \centering
    \begin{tabular}{C{1.5cm}|C{1.5cm}|C{2.25cm}|C{2.25cm}|C{2.25cm}|C{2.25cm}|C{2.25cm}}
        \hline
        \hline
        &  \multirow{2}{*}{$\Delta_{E(\widetilde{E})}$} & \multicolumn{5}{C{11.25cm}}{$L\;\left[\text{fm}\right]$}\\
        \cline{3-7}
        & & 8 & 10 & 12 & 14 & 16\\
        \hline
         \multirow{3}{*}{\rotatebox[origin=c]{90}{$a^{-1}\;\left[\text{fm}^{-1}\right]$}} & $1\%$ & $-0.043_{-0.004}^{+0.004}$ &	$-0.043_{-0.003}^{+0.004}$ &	$-0.043_{-0.003}^{+0.003}$ &	$-0.043_{-0.003}^{+0.003}$ &	$-0.043_{-0.002}^{+0.002}$ \\
        & $5\%$ & $-0.044_{-0.023}^{+0.022}$ &	$-0.044_{-0.017}^{+0.018}$ &	$-0.043_{-0.015}^{+0.014}$ &	$-0.043_{-0.014}^{+0.012}$ &	$-0.044_{-0.011}^{+0.011}$ \\
        & $10\%$ & $-0.049_{-0.052}^{+0.044}$ &	$-0.047_{-0.039}^{+0.035}$ &	$-0.046_{-0.032}^{+0.029}$ &	$-0.045_{-0.030}^{+0.024}$ &	$-0.045_{-0.024}^{+0.021}$ \\
        \hline
        \multirow{3}{*}{\rotatebox[origin=c]{90}{$r_0\;\left[\text{fm}\right]$}} & $1\%$ & $2.751_{-0.097}^{+0.097}$ &	$2.751_{-0.103}^{+0.105}$ &	$2.745_{-0.112}^{+0.119}$ &	$2.753_{-0.130}^{+0.125}$ &	$2.743_{-0.136}^{+0.136}$ \\
        & $5\%$ &
        $2.753_{-0.455}^{+0.533}$ &	$2.751_{-0.488}^{+0.562}$ &	$2.721_{-0.545}^{+0.620}$ &	$2.762_{-0.626}^{+0.666}$ &	$2.715_{-0.668}^{+0.696}$ \\
        & $10\%$&
        $2.751_{-0.839}^{+1.228}$ &	$2.744_{-0.934}^{+1.222}$ &	$2.689_{-1.057}^{+1.362}$ &	$2.771_{-1.254}^{+1.407}$ &	$2.670_{-1.346}^{+1.477}$ \\
         \hline
         \multirow{3}{*}{\rotatebox[origin=c]{90}{$\widetilde{a}^{-1}\;\left[\text{fm}^{-1}\right]$}} & $1\%$ & $0.184_{-0.002}^{+0.002}$ &	$0.184_{-0.002}^{+0.002}$ &	$0.184_{-0.001}^{+0.001}$ &	$0.184_{-0.001}^{+0.001}$ &	$0.184_{-0.001}^{+0.001}$ \\
         & $5\%$ & 
        $0.184_{-0.011}^{+0.012}$ &	$0.184_{-0.009}^{+0.008}$ &	$0.184_{-0.007}^{+0.007}$ &	$0.184_{-0.006}^{+0.006}$ &	$0.184_{-0.005}^{+0.005}$ \\
        & $10\%$ & 
        $0.182_{-0.024}^{+0.023}$ &	$0.183_{-0.018}^{+0.017}$ &	$0.184_{-0.014}^{+0.014}$ &	$0.183_{-0.011}^{+0.012}$ &	$0.184_{-0.010}^{+0.012}$ \\
        \hline
        \multirow{3}{*}{\rotatebox[origin=c]{90}{$\widetilde{r}_0\;\left[\text{fm}\right]$}} & $1\%$ & 
        $1.750_{-0.030}^{+0.028}$ &	$1.749_{-0.030}^{+0.031}$ &	$1.748_{-0.033}^{+0.033}$ &	$1.751_{-0.037}^{+0.034}$ &	$1.747_{-0.037}^{+0.038}$ \\
        & $5\%$ &
        $1.751_{-0.150}^{+0.145}$ &	$1.746_{-0.154}^{+0.150}$ &	$1.740_{-0.171}^{+0.160}$ &	$1.752_{-0.189}^{+0.161}$ &	$1.732_{-0.192}^{+0.187}$ \\
        & $10\%$ &
        $1.747_{-0.303}^{+0.296}$ &	$1.739_{-0.318}^{+0.299}$ &	$1.722_{-0.364}^{+0.318}$ &	$1.745_{-0.403}^{+0.313}$ &	$1.710_{-0.418}^{+0.355}$ \\
         \hline
         \hline
    \end{tabular}
    \caption{Numerical values associated with Fig. \ref{fig: band ERE}.}
    \label{tab: ERE}
\end{table}
\begin{table}[hbt!]
    \renewcommand{\arraystretch}{1.75}
    \centering
    \begin{tabular}{C{0.75cm}|C{1cm}|C{1.5cm}|C{2.25cm}|C{2.25cm}|C{2.25cm}|C{2.25cm}|C{2.25cm}}
        \hline
        \hline
        &  \multirow{2}{*}{$\Delta_\beta$} & \multirow{2}{*}{$\Delta_{E(\widetilde{E})}$} & \multicolumn{5}{C{11.25cm}}{$L\;\left[\text{fm}\right]$}\\
        \cline{4-8}
        & & & 8 & 10 & 12 & 14 & 16\\
        \hline
        \multirow{12}{*}{\rotatebox[origin=c]{90}{$\widetilde{L}_{1,A}\;\left[\text{MeV}\right]$}} &  \multirow{2}{*}{$1\%$}  & \multirow{2}{*}{$1\%$}
        & $-449.4_{-47.0}^{+47.9}$	& $-449.3_{-59.5}^{+70.1}$	& $-447.7_{-82.8}^{+90.5}$	& $-452.4_{-112.6}^{+122.3}$	& $-446.6_{-153.7}^{+134.0}$\\
         & & & $-432.3_{-274.0}^{+208.0}$	& $-451.8_{-136.0}^{+133.8}$	& $-445.1_{-107.1}^{+88.2}$	& $-446.4_{-84.2}^{+71.6}$	& $-450.6_{-68.9}^{+69.3}$\\
         \cline{2-8}
         &  \multirow{2}{*}{$1\%$}  & \multirow{2}{*}{$5\%$}
         & $-451.3_{-135.0}^{+143.0}$	& $-448.0_{-175.6}^{+178.1}$	& $-433.9_{-207.3}^{+205.8}$	& $-441.7_{-257.3}^{+257.3}$	& $-434.5_{-263.4}^{+242.6}$\\
         & & & $-455.4_{-413.1}^{+366.4}$	& $-445.4_{-273.1}^{+223.7}$	& $-444.2_{-206.7}^{+195.2}$	& $-444.5_{-182.8}^{+173.9}$	& $-440.9_{-170.8}^{+153.7}$\\
         \cline{2-8}
         &  \multirow{2}{*}{$1\%$}  & \multirow{2}{*}{$10\%$}
        & $-454.9_{-262.6}^{+273.4}$	& $-456.9_{-324.7}^{+354.6}$	& $-429.9_{-385.6}^{+404.6}$	& $-435.5_{-457.6}^{+465.0}$	& $-415.1_{-470.1}^{+437.5}$ \\
        & & & $-444.0_{-802.8}^{+579.0}$	& $-455.1_{-462.3}^{+407.5}$	& $-447.2_{-376.1}^{+350.8}$	& $-438.4_{-335.8}^{+323.5}$	& $-428.8_{-323.0}^{+286.1}$\\
         \cline{2-8}
         &  \multirow{2}{*}{$5\%$}  & \multirow{2}{*}{$1\%$}& $-439.4_{-195.4}^{+196.8}$	& $-441.5_{-274.8}^{+300.0}$	& $-458.7_{-359.3}^{+447.7}$	& $-466.8_{-541.2}^{+562.6}$	& $-474.5_{-720.3}^{+703.8}$ \\
         & & & $-403.0_{-1952.7}^{+764.7}$	& $-473.3_{-966.3}^{+532.3}$	& $-434.4_{-630.7}^{+358.1}$	& $-440.1_{-527.2}^{+292.6}$	& $-457.3_{-389.6}^{+259.7}$\\
         \cline{2-8}
         &  \multirow{2}{*}{$5\%$}  & \multirow{2}{*}{$5\%$}
        & $-449.2_{-231.1}^{+241.0}$	& $-456.3_{-283.3}^{+363.6}$	& $-450.3_{-397.9}^{+477.0}$	& $-473.6_{-538.8}^{+634.4}$	& $-444.0_{-750.4}^{+681.4}$ \\
        & & & $-387.0_{-1963.6}^{+774.5}$	& $-470.1_{-1039.2}^{+561.2}$	& $-430.1_{-666.8}^{+390.7}$	& $-435.1_{-517.4}^{+316.5}$	& $-467.0_{-397.2}^{+320.6}$\\
         \cline{2-8}
         &  \multirow{2}{*}{$10\%$}  & \multirow{2}{*}{$10\%$}
        & $-466.5_{-433.3}^{+505.0}$	& $-471.0_{-538.4}^{+773.6}$	& $-469.7_{-770.4}^{+991.4}$	& $-522.2_{-1043.3}^{+1300.8}$	& $-474.3_{-1423.8}^{+1407.7}$\\
        & & & $-345.5_{-2191.0}^{+1207.2}$	& $-477.2_{-1981.7}^{+916.3}$	& $-417.7_{-2036.4}^{+690.5}$	& $-423.6_{-1705.4}^{+568.6}$	& $-487.1_{-1167.7}^{+573.3}$\\
        \hline
        \hline
    \end{tabular}
    \caption{Numerical values associated with Fig. \ref{fig: band L1A}. The top (bottom) value in each cell corresponds to the ground-state to ground-state (first excited-state to first excited-state) transition.}
    \label{tab: L1A}
\end{table}
\begin{table}[hbt!]
    \renewcommand{\arraystretch}{1.5}
    \centering
    \begin{tabular}{C{1cm}|C{2.5cm}|C{2cm}|C{2.5cm}|C{2cm}|C{2.5cm}}
        \hline
        \hline
         $L$ & $J^{\infty}(E_0,E_0;m_\pi)$ & $\delta J^V(E_0,E_0)$ & $\big| \mathcal{M}^{\rm (Int.)}_{nn\to pp}\big|$ & $\big| \mathcal{M}^{0\nu,V}_{nn\to pp}\big|$& $\big|\mathcal{T}_L^{(M)}\big|$\\
         $\left[\text{fm}\right]$ & $\left[\text{MeV}^2\right]$ & $\left[\text{MeV}^2\right]$ & $\left[\text{MeV}^{-1}\right]$ & $\left[\text{MeV}^{-1}\right]$ & $\left[\text{MeV}^5\right]$ \\
         \hline
        8 & $9.84\times 10^3$ & $-8.9\times 10^2$ & $2.2\times 10^{-3}$ & $1.95\times 10^{-3}$ & $1.5\times 10^5$\\
        10 & $1.13\times 10^4$ & $-7.5\times 10^2$ & $4.1\times 10^{-3}$ & $3.77\times 10^{-3}$ & $7.6\times 10^4 $\\
        12 & $1.25\times 10^4$ & $-6.3\times 10^2$ & $6.5\times 10^{-3}$ & $6.11\times 10^{-3}$ & $4.2\times 10^4$ \\
        14 & $1.34\times 10^4$ & $-5.3\times 10^2$ & $9.4\times 10^{-3}$ & $8.93\times 10^{-3}$ & $2.5\times 10^4$ \\
        16 & $1.43\times 10^4$ & $-4.4\times 10^2$ & $1.3\times 10^{-2}$ & $1.22\times 10^{-2}$ & $1.5\times 10^4$ \\
         \hline
         \hline
    \end{tabular}
    \caption{Numerical values of the finite- and infinite-volume quantities in the matching relation for the $0\nu\beta\beta$ process in Eq. \eqref{eq: matching relation gvNN}. These quantities are evaluated at the ground-state FV energy eigenvalues in the corresponding volumes.}
    \label{tab: 0vbb Table}
\end{table}
\begin{table}[]
    \renewcommand{\arraystretch}{1.75}
    \centering
    \begin{tabular}{C{1cm}|C{1cm}|C{1.25cm}|C{1.5cm}|C{1.5cm}|C{1.5cm}|C{1.5cm}|C{1.5cm}}
        \hline
        \hline
        &  \multirow{2}{*}{$\Delta_{\beta\beta}$} & \multirow{2}{*}{$\Delta_{E}$} & \multicolumn{5}{C{7.5cm}}{$L\;\left[\text{fm}\right]$}\\
        \cline{4-8}
        & & & 8 & 10 & 12 & 14 & 16\\
        \hline
         \multirow{6}{*}{\rotatebox[origin=c]{90}{$\widetilde{g}_{\nu}^{NN}$}} &  $1\%$  & $1\%$
        & $1.3_{-0.1}^{+0.1}$
        & $1.3_{-0.1}^{+0.1}$
        & $1.3_{-0.1}^{+0.1}$
        & $1.3_{-0.1}^{+0.1}$
        & $1.3_{-0.1}^{+0.1}$\\
         \cline{2-8}
        &  $1\%$  & $5\%$
        & $1.3_{-0.2}^{+0.2}$
        & $1.3_{-0.2}^{+0.2}$
        & $1.3_{-0.3}^{+0.2}$
        & $1.3_{-0.3}^{+0.3}$
        & $1.3_{-0.3}^{+0.3}$\\
        \cline{2-8}
        &  $1\%$  & $10\%$
        & $1.3_{-0.4}^{+0.3}$
        & $1.3_{-0.4}^{+0.4}$
        & $1.3_{-0.5}^{+0.5}$
        & $1.3_{-0.6}^{+0.6}$
        & $1.2_{-0.6}^{+0.6}$\\
         \cline{2-8}
        &  $5\%$  & $1\%$
        & $1.3_{-0.2}^{+0.2}$
        & $1.3_{-0.2}^{+0.2}$
        & $1.3_{-0.2}^{+0.2}$
        & $1.3_{-0.3}^{+0.3}$
        & $1.3_{-0.3}^{+0.3}$\\
         \cline{2-8}
         &  $5\%$  & $5\%$
        & $1.3_{-0.3}^{+0.3}$
        & $1.3_{-0.3}^{+0.3}$
        & $1.3_{-0.3}^{+0.3}$
        & $1.3_{-0.4}^{+0.4}$
        & $1.2_{-0.4}^{+0.4}$\\
         \cline{2-8}
        &  $10\%$  & $10\%$
        & $1.3_{-0.6}^{+0.5}$
        & $1.3_{-0.7}^{+0.6}$
        & $1.2_{-0.7}^{+0.7}$
        & $1.2_{-0.9}^{+0.7}$
        & $1.2_{-0.9}^{+0.8}$\\
        \hline
        \hline
    \end{tabular}
    \caption{Numerical values associated with Fig. \ref{fig: band gvNN}.}
    \label{tab: gvNN}
\end{table}
\bibliography{biblio.bib}

\end{document}